\documentclass{aa}  

\usepackage{graphicx}
\usepackage{caption}
\usepackage{txfonts}
\usepackage{xcolor}
\usepackage{booktabs}
\usepackage{siunitx}
\usepackage{textcomp}

\begin{document}

   \title{The WISSH quasars project XI. The mean Spectral Energy Distribution and Bolometric Corrections of the most luminous quasars}

\titlerunning{The mean SED of the most luminous QSOs}

   \author{Saccheo, I. \inst{1,2},
          Bongiorno, A. \inst{2},
          Piconcelli, E.  \inst{2},
          Testa, V.  \inst{2},
          Bischetti, M. \inst{3,4},
          Bisogni, S. \inst{5},
          Bruni, G. \inst{6},
          Cresci, G.  \inst{7},
          Feruglio, C. \inst{3},
          Fiore, F. \inst{3},
          Grazian, A. \inst{8},
          Luminari, A.  \inst{2,6},
          Lusso, E. \inst{9,7},
          Mainieri, V. \inst{10},
          Maiolino, R. \inst{11,12,13},
          Marconi, A. \inst{7,9},
          Ricci, F.  \inst{1},
          Tombesi, F. \inst{14},
          Travascio, A. \inst{15},
          Vietri, G.  \inst{2},
          Vignali, C. \inst{16,17},
          Zappacosta, L.  \inst{2}
          \and
          La Franca, F.  \inst{1}
          }

   \institute{Dipartimento di Matematica e Fisica, Universitá Roma Tre, Via della Vasca Navale 84, 00146, Roma, Italy;
   \email{ivano.saccheo@uniroma3.it}
   \and 
   INAF, Osservatorio Astronomico di Roma, Via Frascati 33, I– 00078 Monte Porzio Catone, Italy
   \and
   INAF,  Osservatorio Astronomico di Trieste, Via G. B. Tiepolo 11, I–34143 Trieste, Italy
   \and 
   Dipartimento di Fisica, Sezione di Astronomia, Universit\'a di Trieste, via Tiepolo 11, 34143 Trieste, Italy
   \and 
   INAF - Istituto di Astrofisica Spaziale e Fisica cosmica Milano, Via Alfonso Corti 12, 20133, Milano, Italy
   \and
   INAF - Istituto di Astrofisica e Planetologia Spaziali, via del Fosso del Cavaliere 100, 00133 Rome, Italy
   \and 
   INAF – Osservatorio Astrofisco di Arcetri, Largo E. Fermi 5, 50127 Firenze, Italy
   \and
   INAF–Osservatorio Astronomico di Padova, Vicolo dell’Osservatorio 5, I-35122, Padova, Italy
   \and
   Dipartimento di Fisica e Astronomia, Universitá di Firenze, Via G. Sansone 1, 50019, Sesto Fiorentino (Firenze), Italy
   \and
   ESO, Karl-Schwarschild-Strasse 2, D–85748 Garching bei M\"unchen, Germany
   \and
   Kavli Institute for Cosmology, University of Cambridge, Cambridge CB3 0HE, UK
   \and
   Cavendish Laboratory, University of Cambridge, Cambridge CB3 0HE, UK
   \and
   Department of Physics and Astronomy, University College London, Gower Street, London WC1E 6BT, UK
   \and
   Department of Physics, University of Rome ‘Tor Vergata’, Via della Ricerca Scientifica 1, I-00133 Rome, Italy
   \and
   Dipartimento di Fisica ’G. Occhialini’, Università degli Studi di Milano-Bicocca, Piazza della Scienza 3, I-20126 Milano, Italy
   \and
    Dipartimento di Fisica e Astronomia “Augusto Righi”, Università degli Studi di Bologna, via P. Gobetti 93/2, 40129 Bologna, Italy
   \and
   INAF - Osservatorio di Astrofisica e Scienza dello Spazio, via P. Gobetti 93/3, 40129, Bologna, Italy
   }
   
\titlerunning{The mean SED of the most luminous QSOs}
\authorrunning{Saccheo et al.}
   \date{Received June 20, 2022; Accepted November 10, 2022}

 
  \abstract
   {Hyper-luminous Quasi-Stellar Objects (QSOs)
  represent the ideal laboratory to investigate Active Galactic Nuclei (AGN) feedback mechanism  since their formidable energy release causes powerful winds at all scales and thus the maximum feedback is expected.} 
   {We aim at deriving the mean Spectral Energy Distribution (SED) of a sample of 85 WISE-SDSS Selected Hyper-luminous (WISSH) quasars. Since the  SED provides a direct way to investigate the AGN structure, our goal is to understand if quasars at the bright end of the luminosity function have peculiar properties compared to the bulk of the QSO population.}
   {We collected all the available photometry, from X-ray to FIR:  each WISSH quasar is observed in at least 12 different bands. We  then built a mean intrinsic SED after correcting for the dust extinction, absorption and emission lines, and intergalactic medium absorption. We also derived bolometric, IR band and monochromatic luminosities together with bolometric corrections at $\lambda = 5100\,\text{\AA}$ and $3$ \textmu m. We define a new relation for the $3$ \textmu m bolometric correction.}
   {We find that the mean SED of hyper-luminous WISSH QSOs shows some differences compared to that of less luminous sources, i.e. a relatively lower X-ray emission and a near and mid IR excess which can be explained assuming a larger dust contribution. WISSH QSOs have stronger emission from both warm  ($T\sim 500-600$ K) and very hot ($T\geq 1000$ K) dust, the latter being responsible for shifting the typical dip of the AGN SED from $1.3$ \textmu m to $1.1$ \textmu m. We also derived the mean SEDs of two sub-samples created based on their spectral features (presence of Broad Absorption Lines  and equivalent width of CIV line). We confirm that BALs are X-ray weak and that they have a reddened UV-optical continuum. We also find that BALs tend to have stronger emission from the hot dust component. For what concerns sources with a weaker CIV line, our main result is the confirmation of their lower X-ray emission. By populating the $L_{IR}$ vs $z$ diagram proposed by \cite{Symeonidis:2021}, we found that $\sim$ 90\% of WISSH QSOs with $z\geq3.5$ have their FIR emission dominated by star forming activity.}
   {This analysis suggests that hyper-luminous QSOs have a peculiar SED compared to less luminous objects. It is therefore critical to use SED templates constructed exclusively  from very bright quasars samples (such as this)  when dealing with particularly luminous sources, such as high resdhift QSOs.}


  \maketitle
%

\section{Introduction}
The Spectral Energy Distribution (SED) describes the  emission of a source throughout  the electromagnetic spectrum and is therefore a powerful tool to investigate the physical processes originating it. In the case of Active Galactic Nuclei (AGN), the study of their SED is particularly interesting since it extends to the whole electromagnetic spectrum, from the hard X-rays to the radio band. \\
Indeed, the AGN SED is due to the sum of several contributions, arising from distinct regions and from different physical mechanisms.
 In particular, the actively accreting supermassive black hole (SMBH) is powered by the inflow of gas through an accretion disk with the subsequent transformation of gravitational energy into thermal energy due to viscous torques \citep[][]{Lynden-Bell:1969}. This process is responsible of the prominent 'blue' bump in the UV and optical emission. The primary radiation can then be Compton up-scattered to X-ray energies  in a region consisting of  hot electron gas called corona \citep[e.g.][]{Liang:1979}. Moreover, a dusty torus, surrounding the accretion disk, absorbs UV and optical photons and re-emits them in the near and mid IR. In a similar way, dust at a lower temperature ($T\approx 20-100$ K) located at much greater distances, heated by hot stars and partly by the central AGN, is responsible for the emission in the far IR.
Finally, the possible presence of a relativistic jet explains the radio emission in about 10\% of the AGN \citep[e.g][]{Blandford:1982}. The study of SEDs is therefore primarily a direct way to investigate the AGN structure and a powerful tool to better understand the physical phenomena which are taking place.\\
Moreover, a precise determination of the SED allows us to compute the overall energy output of the AGN 
given by the bolometric luminosity $L_{bol}$, i.e. the integral under the SED.
This is an important observational parameter in several studies and models of AGN feedback and AGN-host galaxy co-evolution \citep[e.g.,][]{Ishibashi:2012}. 
Finally,  the knowledge of the SED allows to derive the Bolometric Corrections, i.e. the ratio between $L_{bol}$ and the luminosity in a given band, which are necessary to estimate the bolometric luminosity of  sources for which multi-wavelength observations are not available.\\
\cite{Elvis:1994} firstly gave the composite SED of a sample of 47 Type 1 quasars  from the PG catalogue extending from the X-rays to the radio band. In the following years SEDs from an increasing number of QSOs were derived \citep[e.g.,][]{Hatziminaoglou:2005,Shang:2011, Bianchini:2019a}. Notably, \cite{Richards:2006} computed the mean SED from a sample of 259 quasars with both \textit{SDSS} and \textit{Spitzer IRAC } photometry. \cite{Krawczyk:2013} (hereafter K13)  extended their work  using a sample of 108,184 non-reddened  type 1 quasars at $0.064 < z < 5.46$ with at least \textit{SDSS} photometry   to construct a mean SED between $10$ keV and $\sim 20.0$ \textmu m which is the most robust according to the number of sources used.\\
Both \cite{Richards:2006} and K13 constructed the composite SEDs of some subsamples based on the luminosity of the QSOs; SEDs obtained in these works are similar to the one originally derived by \cite{Elvis:1994} although with small differences,  and their study reveals that the shape of AGN emission is generally comparable over a wide range of luminosities. \cite{Hao:2014}, using 407 quasars from \textit{COSMOS} investigated the dependence of the SED shape on several physical properties of the sources (redshift, $L_{bol}, M_{BH}$, Eddington ratio) and no significant correlation was found.\\
However, both the fact that bolometric corrections seem to be luminosity dependent  \citep[e.g.][]{Runnoe:2012,Duras:2020} and that the ratio between X-ray and optical luminosity (i.e. $\alpha_{OX}$) decreases for optically bright QSOs \citep[e.g][]{Steffen:2006} necessarily implies some kind of evolution in the shape of the SED. Indeed, while SEDs derived from large statistical samples are robust and exceptionally effective in representing the average properties, they might fail in the  description of distinctive features specific to a particular subclass of AGN.  
For this reason, when analyzing sources with some peculiar properties, it might not be appropriate to rely on SEDs built from extremely varied samples, but rather it is preferable to use those that have been derived from QSOs with similar features. \\ 
A population of AGN particularly interesting to study is represented by quasars at the bright end of the luminosity function, since both theory and observations suggest that these are the sources where feedback is stronger \citep[e.g.][]{Veilleux:2013, Cicone:2014, Fiore:2017,Bischetti:2019,Fluetsch:2019,Lutz:2020}. Although in the catalog used by K13 there are several QSOs with $L_{bol}$ up to $10^{48}$ erg/s, these are only a small fraction even considering their sub-sample of optically bright sources red($\sim 2.8\%$ of the whole sample has $L_{bol}\geq10^{47}$ erg/s). Therefore the resulting composite SED is not really representative of these few hyper-luminous sources but of the bulk of the population at lower luminosities.\\
In this Paper we compute the composite SED of WISSH quasars, a sample consisting of 85 type 1 QSOs selected to be among the most luminous in the Universe.\\
Throughout this paper we assume a flat $\Lambda CDM$ cosmology with $H_{0} = 70$ km/s $Mpc^{-1}$ and $\Omega_{\Lambda} = 0.70$.

\section{The WISSH Sample}
The WISSH sample \citep[][]{Bischetti:2017} is composed of 85 type 1, radio-quiet hyper-luminous quasars with $\log(L_{bol}/[\,erg\, s^{-1}\,]) \geq 47.0 $ \citep[][]{Duras:2017}.\\
The sample has been assembled cross-correlating  the  \textit{SDSS} \citep[][]{Shen:2011} and the \textit{WISE} \citep[][]{Wright:2010} catalogues and selecting, among the sources with a flux density $S_{\nu, 22 \mu m}> 3$ mJy, the 100 most luminous QSOs at $\lambda = 7.8$ \textmu m \citep[][]{Weedman:2012}. Gravitationally lensed objects , those that had a contaminated \textit{WISE} photometry and those with an anomalous radio emission have  then been removed, leaving the final sample of 85 quasars with $46.96\leq L_{7.8\mu m}/[erg\,s^{-1}] \leq 47.49$ and a redshift distribution $1.8<z < 4.7$.\\
Given their high luminosities and their redshift distribution covering the epoch where both AGN activity and star formation were at their peak  \citep[e.g.][]{Ueda:2003,Hasinger:2005,Madau:2014} WISSH quasars represent the ideal targets to investigate the feedback mechanism.
Several previous works on the sample have indeed shown the exceptional nature of these sources in their ability to drive powerful winds at different scales. \\
For example, \cite{Bruni:2019}, analyzing the Broad Absorption Lines (BAL)  population in WISSH, found that it represents a larger fraction when compared to other samples (24\% of sources with CIV Balnicity index > 0 compared to 13.5\% found by \citealt{Gibson:2009}), confirming that high $L_{bol}$ favors the acceleration of nuclear scale winds. Moreover, using the [OIII] and CIV emission lines as tracers of ionized gas, the velocity and the kinetic power of the outflows have been measured for different sources on scales up to 7-10 kpc, finding mass outflow rates among the highest reported in the literature \citep[][]{Bischetti:2017, Vietri:2018}. Also, from the analysis of Lyman-$\alpha$ emitting nebulae surrounding one QSO, \cite{Travascio:2020} found evidence for outflowing gas at  circum-galactic scale.
Finally, it has been found that these quasars are located in extremely overdense environments \citep[see][]{Bischetti:2018,Bischetti:2021}.

\begin{table*}
\centering
\begin{tabular}[t]{cccS[table-format=1.4]||}
ID  & RA & DEC & z \\
\hline
\hline
WISSH01                                                       & 00 45 27.68                              & +14 38 16.1                               & 1.9897$^{d}$  \\
WISSH02                                                  & 01 24 03.77                              & +00 44 32.6                               & 3.822$^{d}$                                   \\
WISSH03                                     & 01 25 30.85                              & -10 27 39.8                               & 3.3588$^{d}$                                   \\
WISSH04                                               & 02 09 50.71                              & -00 05 06.4                               & 2.870$^{c}$                                   \\
WISSH05                                                    & 02 16 46.94                              & -09 21 07.2                               & 3.7387$^{d}$                                  \\
WISSH06                                          & 04 14 20.90                              & +06 09 14.2                               & 2.6324$^{e}$                                   \\
WISSH07                                                           & 07 35 02.30                              & +26 59 11.5                               & 1.999$^{d}$                                   \\
WISSH08                                                         & 07 45 21.78                              & +47 34 36.1                               & 3.225$^{a}$                                \\
WISSH09                                                 & 07 47 11.14                              & +27 39 03.3                               & 4.126$^{e}$                                   \\
WISSH10                                          & 08 01 17.79                              & +52 10 34.5                               & 3.257$^{b}$                               \\
WISSH11                                                & 08 18 55.77                              & +09 58 48.0                               & 3.6943$^{f}$                                    \\
WISSH12                                                          & 08 46 31.52                              & +24 11 08.3                               & 4.7218$^{d}$                                   \\
WISSH13                                                        & 09 00 33.50                              & +42 15 47.0                               & 3.294$^{a}$                                   \\
WISSH14                                                        & 09 04 23.37                              & +13 09 20.7                               & 2.9765$^{d}$                                   \\
WISSH15                                                   & 09 28 19.29                              & +53 40 24.1                               & 4.466$^{d}$                                  \\
WISSH16                                                   & 09 41 40.17                              & +32 57 03.2                               & 3.454$^{d}$                                  \\
WISSH17                                                         & 09 47 34.19                              & +14 21 16.9                               & 3.031$^{e}$                                   \\
WISSH18                                                       & 09 50 31.63                              & +43 29 08.4                               & 1.7696$^{d}$                                   \\     
WISSH19                                                      & 09 58 41.21                              & +28 27 29.5                               & 3.434$^{b}$                                 \\
WISSH20                                                        & 09 59 37.11                              & +13 12 15.4                               & 4.0781$^{d}$                                 \\
WISSH21                                                      & 10 13 36.37                              & +56 15 36.3                               & 3.6507$^{d}$                                   \\
WISSH22                                                         & 10 14 47.18                              & +43 00 30.1                               & 3.1224$^{d}$                                  \\
WISSH23                                                      & 10 15 49.00                              & +00 20 20.0                               & 4.407$^{c}$                                   \\
WISSH24                                                        & 10 20 40.61                              & +09 22 54.2                               & 3.6584$^{d}$                                   \\
WISSH25                                                           & 10 25 41.78                              & +24 54 24.2                               & 2.3917$^{d}$                                   \\
WISSH26                                                        & 10 26 32.97                              & +03 29 50.6                               & 3.8808$^{d}$        \\
WISSH27                                              & 10 27 14.77                              & +35 43 17.4                               & 3.1182$^{d}$                                  \\
WISSH28                                                           & 10 48 46.63                              & +44 07 10.8                               & 4.408$^{d}$                                  \\
WISSH29                                                         & 10 51 22.46                              & +31 07 49.3                               & 4.2742$^{d}$                                 \\
WISSH30                                                           & 10 57 56.25                              & +45 55 53.0                               & 4.1306$^{d}$                                   \\
WISSH31                                                            & 11 03 52.74                              & +10 04 03.1                               & 3.6004$^{d}$                                    \\
WISSH32                                                         & 11 06 07.47                              & -17 31 13.5                               & 2.572$^{e}$                                   \\
WISSH33                                                   & 11 06 10.72                              & +64 00 09.6                               & 2.221$^{b}$                                   \\
WISSH34                                                         & 11 10 17.13                              & +19 30 12.5                               & 2.502$^{e}$                                  \\
WISSH35                                                      & 11 10 38.63                              & +48 31 15.6                               & 2.9741$^{d}$                                   \\
WISSH36                                                            & 11 10 55.21                              & +43 05 10.0                               & 3.8492$^{d}$                                  \\
WISSH37                                                        & 11 11 19.10                              & +13 36 03.9                               & 3.490$^{b}$                                   \\
WISSH38                                                        & 11 22 58.77                              & +16 45 40.3                               & 3.0398$^{f}$                                  \\
WISSH39                                                       & 11 30 17.37                              & +07 32 12.9                               & 2.659$^{d}$                                  \\
WISSH40                                                   & 11 57 47.99                              & +27 24 59.6                               & 2.217$^{b}$                                   \\
WISSH41                                                        & 11 59 06.52                              & +13 37 37.7                               & 4.0084$^{d}$                                 \\
WISSH42                                                        & 12 00 06.25                              & +31 26 30.8                               & 2.9947$^{d}$                                  \\
WISSH43                                                         & 12 01 44.36                              & +01 16 11.6                               & 3.248$^{b}$                                \\
\end{tabular}
\begin{tabular}[t]{cccS[table-format=1.4]}
ID  & RA & DEC & z \\
\hline
\hline

WISSH44                                                          & 12 01 47.90                              & +12 06 30.2                               & 3.512$^{a}$                                  \\
WISSH45                                                   & 12 04 47.15                              & +33 09 38.7                               & 3.638$^{d}$                                 \\
WISSH46                                                     & 12 10 27.62                              & +17 41 08.9                               & 3.831$^{g}$                                  \\
WISSH47                                                   & 12 15 49.81                              & -00 34 32.1                               & 2.6987$^{d}$                                 \\
WISSH48                                                          & 12 19 30.77                              & +49 40 52.2                               & 2.6928$^{d}$                                   \\
WISSH49                                                         & 12 20 16.87                              & +11 26 28.1                               & 1.8962$^{d}$                                  \\
WISSH50                                                          & 12 36 41.45                              & +65 54 42.1                               & 3.424$^{b}$                                 \\
WISSH51                                               & 12 37 14.60                              & +06 47 59.5                               & 2.7891$^{d}$                                  \\
WISSH52                                                         & 12 45 51.44                              & +01 05 05.0                               & 2.8068$^{d}$                                  \\
WISSH53                                                       & 12 49 57.23                              & -01 59 28.8                               & 3.6286$^{d}$                                   \\
WISSH54                                                        & 12 50 05.72                              & +26 31 07.5                               & 2.0476$^{e}$                                   \\
WISSH55                                                    & 12 50 50.88                              & +20 46 58.7                               & 3.543$^{e}$                                   \\
WISSH56                                             & 13 05 02.28                              & +05 21 51.1                               & 4.101$^{d}$                                 \\
WISSH57                                            & 13 10 11.60                              & +46 01 24.4                               & 2.1423$^{d}$                                 \\
WISSH58                                                  & 13 26 54.96                              & -00 05 30.1                               & 3.303$^{a}$                                   \\
WISSH59                                                     & 13 28 27.06                              & +58 18 36.8                               & 3.140$^{d}$                                  \\
WISSH60                                                         & 13 33 35.78                              & +16 49 03.9                               & 2.099$^{h}$                                   \\
WISSH61                                                        & 14 21 23.97                              & +46 33 18.0                               & 3.454$^{b}$                                 \\
WISSH62                                                       & 14 22 43.02                              & +44 17 21.2                               & 3.648$^{b}$                                  \\
WISSH63                                                        & 14 26 56.18                              & +60 25 50.8                               & 3.1972$^{d}$                                   \\
WISSH64                                                   & 14 33 52.21                              & +02 27 13.9                               & 4.728$^{c}$                                   \\
WISSH65                                               & 14 41 05.53                              & +04 54 54.9                               & 2.080$^{h}$                                  \\
WISSH66                                                    & 14 47 09.24                              & +10 38 24.5                               & 3.7042$^{d}$                                  \\   
WISSH67                                                          & 14 51 25.31                              & +14 41 36.0                               & 3.094$^{e}$                                  \\    
WISSH68                                                  & 15 06 54.55                              & +52 20 04.7                               & 4.0995$^{d}$                                  \\    
WISSH69                                                       & 15 13 52.52                              & +08 55 55.7                               & 2.8883$^{d}$                                   \\   
WISSH70                                                    & 15 21 56.48                              & +52 02 38.5                               & 2.218$^{b}$                                \\            
WISSH71                                                  & 15 38 30.55                              & +08 55 17.0                               & 3.567$^{b}$                                   \\
WISSH72                                                        & 15 44 46.34                              & +41 20 35.7                               & 3.5513$^{d}$                                  \\
WISSH73                                            & 15 49 38.72                              & +12 45 09.1                               & 2.365$^{a}$                                  \\
WISSH74                                                        & 15 54 34.17                              & +11 09 50.6                               & 2.930$^{e}$                                   \\
WISSH75                                                       & 15 55 14.85                              & +10 03 51.3                               & 3.529$^{c}$                                 \\
WISSH76                                                      & 15 59 12.34                              & +48 28 19.9                               & 3.419$^{d}$                                   \\
WISSH77                                                          & 15 59 52.67                              & +19 23 10.4                               & 3.9532$^{f}$                                   \\
WISSH78                                                          & 16 21 16.92                              & -00 42 50.8                               & 3.7285$^{d}$                                  \\
WISSH79                                                        & 16 33 00.13                              & +36 29 04.8                               & 3.5747$^{d}$                                   \\
WISSH80                                                      & 16 39 09.10                              & +28 24 47.1                               & 3.846$^{c}$                                   \\
WISSH81                                                          & 16 50 53.78                              & +25 07 55.4                               & 3.337$^{e}$                                   \\
WISSH82                                                      & 17 01 00.60                              & +64 12 09.3                               & 2.753$^{c}$                                  \\
WISSH83                                                         & 21 23 29.46                              & -00 50 52.9                               & 2.282$^{b}$                                 \\
WISSH84                                                       & 22 38 08.07                              & -08 08 42.1                               & 3.1422$^{d}$                                   \\
WISSH85                                                        & 23 46 25.66                              & -00 16 00.4                               & 3.511$^{b}$                               \\

\end{tabular}

\caption{The WISSH sample: ID, Coordinates and Redshift. The source name is SDSSJ followed by RA and DEC (e.g. \textit{SDSSJ004527.68+143816.1} for WISSH01)\\
Redshift provided by  $^{a}$\cite{Bischetti:2017}, $^{b}$\cite{Vietri:2018}, $^{c}$\cite{Bischetti:2021}, $^{d}$\cite{Hewett:2010}, $^{e}$ \cite{Paris:2014}, $^{f}$ primary z by \cite{Lyke:2020},  $^{g}$\cite{Yi:2020}, $^{h}$\cite{Vietri:2020}.}
\label{tab:targets}
\end{table*}

In table \ref{tab:targets} WISSH quasars  are listed along with their coordinates, their redshifts and their identification number. Redshifts, when available, are taken from \cite{Bischetti:2017, Vietri:2018, Bischetti:2021,Yi:2020} and form the SUPER survey \citep[][]{Kakkad:2020,Vietri:2020}; otherwise we use values provided by \cite{Hewett:2010} who derived redshifts without including the CIV emission line in the computation; indeed as shown by \cite{Vietri:2018} the CIV line can lead to a systemic  underestimation of the redshift. This leaves with 12 QSOs without an assigned redshift; for these sources we adopt the value provided by the most recent SDSS Quasar Catalog Data Release \citep[][]{Lyke:2020, Paris:2014}.

\section{Multi-wavelength photometry}
\begin{table*}
\begin{center}
\begin{tabular}{ll}

Column & Description\\
\hline
\hline
1 & ID of the sources\\
2-11 & \textit{SDSS ugriz} filters\\
12-17 &  \textit{J,H,K} bands from \textit{2MASS} or \textit{TNG} (with a '-' sign)\\
18-25 &  $3.3\,\mu m, \;4.6\, \mu m, \;12 \,\mu m, \;22 \,\mu m$  filters from \textit{WISE}\\
26-31 & $250\,\mu m,\; 350\, \mu m,\; 500 \,\mu m $  filters from \textit{Herschel}\\
32 & \textit{ALMA} or \textit{NOEMA} observed band [GHz]\\
33-34 & \textit{ALMA} or \textit{NOEMA} observed flux [mJy]\\
35 & \textit{JVLA} observed band [GHz]\\
36-37 & \textit{JVLA} observed flux [mJy]\\
38 & \textit{Chandra} or \textit{XMM} 2-10 keV flux [$10^{-14}\, erg\,s^{-1} \,cm^{-2} $]\\
39 & X-Ray Spectral index $\Gamma$ \\
\hline
\end{tabular}

\end{center}

\caption{Reference table for the WISSH sample photometry. The full table is available in the online version. 
Unless otherwise specified, values are in AB magnitudes}
\label{tab:magnitudini_lista}
\end{table*}

By construction of the sample, all 85 WISSH QSOs have \textit{SDSS} photometry in the \textit{ugriz} filters and have been detected in each of the 4 \textit{WISE} bands at $\lambda = 3.3, \; 4.6,\; 12 $ and $22$ \textmu m. In addition to that, \textit{2MASS} photometry in the J, H and K bands has been collected for roughly 80\% of the sample while the remaining QSOs were targets of an observational campaign conducted by our group with \textit{TNG-NICS} (see Appendix \ref{sec:TNG_data}). Therefore, for all sources we have photometry in 12 observed bands between $\lambda = 3500\,\text{\AA}$ and $\lambda =22$ \textmu m.\\
Moreover, in the far IR, \cite{Duras:2017} used \textit{Herschel} archival data to recover the flux density of 16 QSOs for the three SPIRE bands at $\lambda = 250,\;350$ and $500$ \textmu m. The far IR coverage is also provided by the   observations of 9 QSOs (5 of which are among those with \textit{Herschel} data) performed with \textit{ALMA}, \textit{NOEMA} and \textit{JVLA} in the $\sim 30 -350$ GHz frequency range  \citep[][]{Bischetti:2021}. \\ 
Given the poor \textit{Herschel} angular resolution (18.1", 25.2" and 36.6" beams {FWHM}, at $\lambda =$ 250, 350 and 500 \textmu m respectively, vs $\sim1.0\times0.8$ arcsec$^{2}$ and $\sim 3.5\times 2.1$ arcsec$^{2}$ of \textit{ALMA} and \textit{NOEMA}, respectively), the measured flux could be contaminated by the emission of nearby sources. \cite{Bussmann:2015, Trakhtenbrot:2017, Hatziminaoglou:2018} estimated the multiplicity rate to be between 30\% and 50\%.  
In the WISSH sample, \textit{ALMA} and \textit{NOEMA} observations revealed the presence of at least one nearby companion galaxy, \citep[i.e. in the same FOV of the telescope, that has a diameter of about 20-30" for both \textit{ALMA} and \textit{NOEMA},][]{Bischetti:2018} in 7 out of 9 QSOs, although only in two cases their fluxes have been constrained and are found to represent 50\% and 27\% of the total observed emission. For the other QSOs, only 3-$\sigma$ upper limits on their neighboring companions contribution could be determined and range between 0.2\% and 54\% \citep[][]{Bischetti:2021}.
For the 5 QSOs with both \textit{Herschel} and \textit{ALMA/NOEMA} data, we subtracted companion fluxes in the 250, 350 and 500 \textmu m bands assuming that they made the same contribution as observed with \textit{ALMA/NOEMA} (for those sources with unconstrained fluxes, we removed half of their upper limits). For the remaining \textit{Herschel}-observed sources we did not apply any corrections but, based on what we found for the other QSOs, we estimate that, on average, their fluxes may be overestimated by roughly $\sim25\%$.

Finally, in the X-ray region, the $2-10$ keV integrated fluxes of 43 sources are available thanks to \textit{Chandra} or \textit{XMM-Newton} detections \citep[][]{Martocchia:2017, Zappacosta:2020}. \\
Although WISSH photometry has been collected over an extended period of time and the same QSOs might have been observed in different bands a long time apart, we do not expect our data to be drastically affected by source variability. Indeed it has been shown  \citep[e.g.][]{Vanden-Berk:2004, Caplar:2017} that AGN variability clearly anti-correlates with luminosity and therefore it should not constitute a real issue for WISSH hyper-luminous sources.
A table with the entire photometry of the WISSH sample is reported in the online version of the journal. The description of the columns is given in table \ref{tab:magnitudini_lista}. Magnitudes are AB and have been corrected for the galactic extinction \citep[][]{Schlafly:2011}.

\section{Construction of the mean SED}

\subsection{Removing photometric data significantly affected by extinction}
\label{sec:removing_red}

The analyzed sample is composed of type 1 unobcured QSOs. 
Nevertheless, a fraction of them shows a redder spectrum compared to classical, unobscured quasars, and it can be explained in terms  of dust reddening of the (bluer) intrinsic spectrum.\\
To select obscured quasars  we computed the optical spectral slope $\alpha_{opt}$ between $\lambda = 0.3$ \textmu m and $\lambda = 1$ \textmu m and identified as obscured those with $\alpha_{opt} < 0.2$, corresponding to an $E_{B-V} \geq 0.15$ assuming the mean SED from \cite{Richards:2006} \citep[see fig. 5 in ][]{Bongiorno:2012}. A total of 7 QSOs (WISSH25, WISSH34, WISSH49, WISSH55, WISSH58, WISSH73 and WISSH74) were found to satisfy this criterion. These QSOs also satisfy the criterion proposed by \cite{Glikman:2004} to identify red quasars, i.e. $r-K > 4$ and $J-K > 1.7$, while only WISSH74 would also be marked as an extremely red QSOs according to the criterion by \cite{Ross:2015}, i.e.  $r-W4 \gtrsim 7.5$. Moreover, \cite{Glikman:2022} define as \textit{red QSOs} sources with $E_{B-V}>0.25$; according to this definition only WISSH34, WISSH49 and WISSH74 are properly red (see tab. \ref{tab:lbol}), for the other 4 sources it would perhaps be more appropriate to use the term \textit{reddish}.
Since the blue part of the  SED of these objects does not reflect their intrinsic emission, we excluded their photometry at wavelengths shorter than $1.0$ \textmu m in the computation of the mean SED.\\
Moreover,  since the work from \cite{Bruni:2019} outlined  the presence of BAL in a significant fraction of the objects, we visually inspected the \textit{SDSS} spectra and removed the photometry that was clearly depressed due to the presence of BAL features. After corrections have been applied (see Sec. \ref{subsec:corrections}), we replaced  these points with non-absorbed data, using the mean SED by K13 normalized to the nearest filter available (\textit{Gap repair}). This procedure was applied for a total of 11 QSOs, mostly for the \textit{u} and \textit{g} bands. Their \textit{repaired} photometry is shown as grey circles in fig. \ref{fig:mean_SED}.

\subsection{Corrections}
\label{subsec:corrections}
There are several factors that modify the observed radiation with respect to the \textit{intrinsic} emission of an AGN. Since we are interested in reconstructing  the mean intrinsic SED of hyper-luminous QSOs, it is necessary to determine the contributions provided by these factors and correct the data for them. In particular, following the same procedure also adopted by K13 we considered  the absorption of the intergalactic medium, the contamination of the spectral emission lines from Broad and Narrow Line Regions and the contribution from the host galaxy. The methods we employed to determine the corrections to be applied are indirect and are based on statistical grounds. To this extent, the correcting factors they provide for a single QSO are not necessarily true; however, when averaged over the whole sample, they should give a fairly accurate result. For such a reason we used corrected photometry only in the derivation of the mean SED(s); when dealing with the computation of individual QSOs properties (e.g. Sec. \ref{sec:bolometric_lum}), we do not apply any corrections.

\subsubsection*{(i) The Intergalactic Medium absorption}
\label{sec:IGM_correction}
One of the main problems that arise when  reconstructing the SED of a distant source is the absorption due to the intergalactic medium (IGM) mainly constituted of neutral Hydrogen clouds distributed along the line of sight and
responsible for the drastic drop of the observed flux blueward the Ly$\alpha$ line at $\lambda = 1216$ \AA. One way to account for this effect is to assume a statistical distribution of the absorbers and evaluate the expected transmission function   $T_{\lambda}=e^{-\tau(z, \lambda)}$ where $\tau$ is the optical depth.\\
We used the results from \cite{Inoue:2014} who developed an analytical model to estimate the optical depth $\tau$ as a function of the source's redshift.
In the model from \cite{Inoue:2014} the evolution of the number density distribution of the HI clouds is described separately for  two components, the \textit{Ly$\alpha$ forest} component and the \textit{Damped Ly$\alpha$ systems}, the former being dominant for a column density $\log(N_{HI}/cm^{-2})<17.2$ and the latter for $\log(N_{HI}/cm^{-2}) \geq 20.3$ and with a mixed contribution for intermediate values. 
To be conservative in our corrections, we assumed no Damped Ly$\alpha$ component when evaluating the optical depth.
Once derived the optical depth, it  can be transformed in magnitude correction for the five \textit{SDSS} filters, by convolving the IGM attenuation with the filter transmission $S_{\lambda}$ and the continuum flux $F_{\lambda}$
\begin{equation*}
\label{eqn:magnitude_correction_extinction}
\Delta m_{IGM} = -2.5 \log\frac{\int \lambda F_{\lambda} e^{-\tau(\lambda ,z)}S_{\lambda}d\lambda}{\int \lambda F_{\lambda} S_{\lambda}d\lambda}
\end{equation*}
where the UV-to-Optical continuum flux for each QSO was modeled as a single power-law $F_{\lambda} \propto \lambda^{-1.56}$  \citep{Vanden-Berk:2001}.
Non-extincted magnitudes are then recovered as:
\begin{equation*}
    m(\tau = 0) = m(\tau) -\Delta m_{IGM}
\end{equation*}
We find that, for the analyzed sample, $\Delta m_{IGM}$ ranges between 0.02 and 3.96 (1.21 on average) for the \textit{u} filter and between 0 and 2.65 (0.48 on average) for the \textit{g} one.

\subsubsection*{(ii) Emission Lines}

\begin{figure}
    \includegraphics[width = 0.5\textwidth]{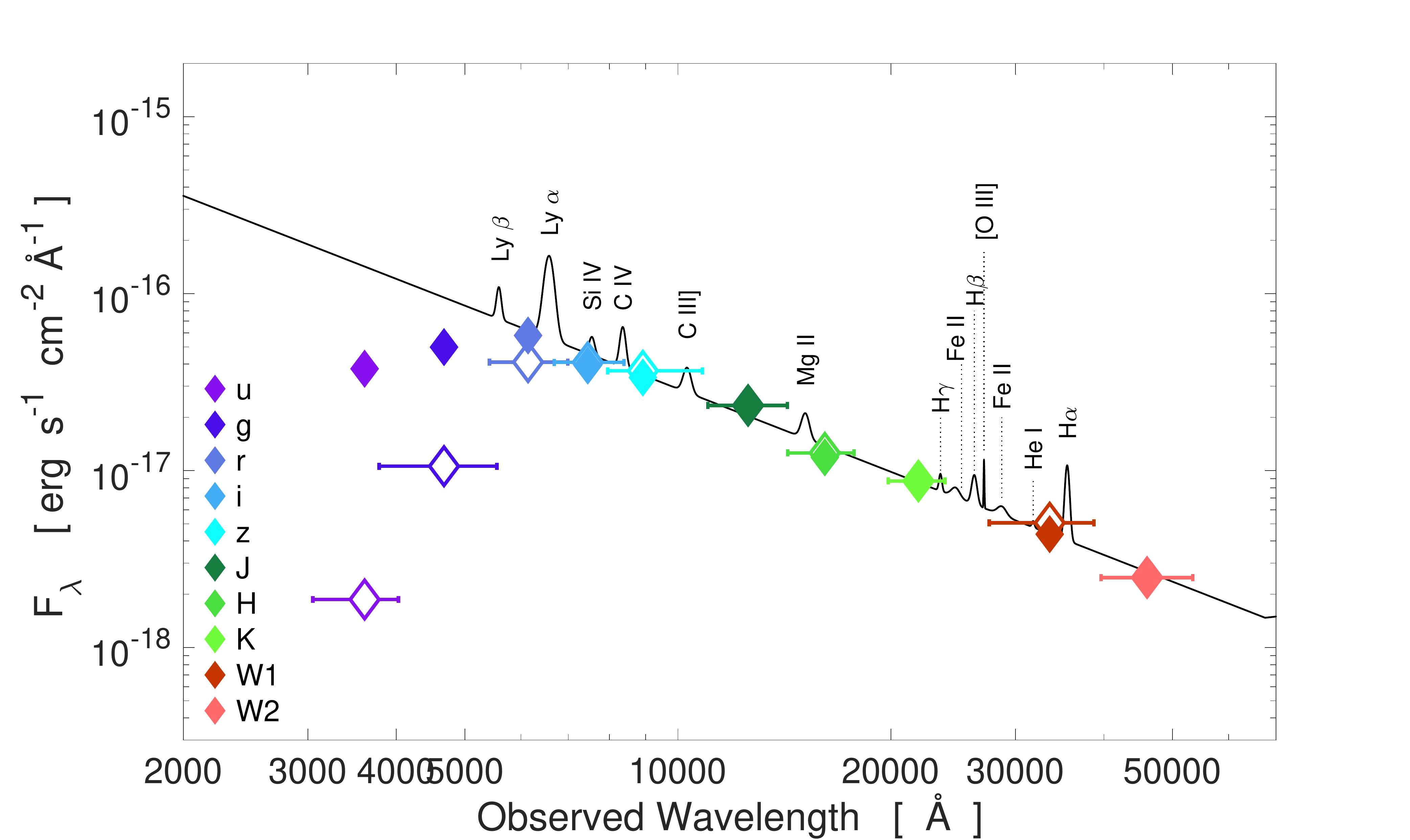}
    \caption{Example of corrections applied to take into account both IGM absorption and emission lines for WISSH23 at $z = 4.4$. The empty diamonds represent the observed fluxes  while the filled diamonds mark the fluxes after corrections. The horizontal bars indicate the range of wavelengths covered by the filters.}
    \label{fig:corrections}
\end{figure}

AGN are characterized by several emission lines in their spectra. If these lines fall into an observed filter, they lead to an overestimation of the source flux. 
In order to remove this effect we built a mock template of the continuum flux $f(c)$ and we added the 13 strongest emission lines measured by \cite{Vanden-Berk:2001} in their composite spectrum, $f(c\&l)$. As in Sec. \ref{sec:IGM_correction}\,(i) the continuum flux was modelled as a single power-law while the emission lines have been modelled as Gaussian functions using the equivalent widths and the FWHMs provided by \cite{Vanden-Berk:2001} (see fig. \ref{fig:corrections}). A possible issue related to this procedure is that we neglected the well known luminosity dependence of the lines equivalent width \citep[e.g.][]{Baldwin:1989,Pogge:1992,Bian:2012}. Moreover, in this  approximation, for  simplicity reasons, the skewness of the line profiles was also neglected.
We also neglected the equivalent width dependence \\
The $f(c\&l)$ template is shifted for each source, depending on its redshift. The correction is:
\begin{equation*}
    \label{eqn:magnitude_correction_lines}
    \Delta m_{EL} = -2.5 \log\frac{\int \lambda F_{\lambda}(c\&l)S_{\lambda}d\lambda}{\int \lambda F_{\lambda}(c) S_{\lambda}d\lambda}
\end{equation*} 
Therefore, the corrected magnitudes are given by:
\begin{equation*}
    m(c) = m(c\&l) - \Delta m_{EL}
\end{equation*}
The correction was performed  over  the five \textit{SDSS} filters, the $J$, $H$ and $K$ filters from \textit{2MASS} or \textit{TNG}   and for W1 and W2 in which emission lines fall given the redshift range of each WISSH QSO. 
Figure \ref{fig:corrections} shows the observed and corrected (both for IGM and emission lines) fluxes for a source at $z=4.40$ as an example.

\subsubsection*{(iii) Host Galaxy contamination}
\label{sec:host_correction}
Another contribution  that should be removed before computing the intrinsic AGN SED is provided by the host galaxy. In the absence of observations allowing for a direct measure of the light coming from the host, we must infer $L_{host}$ from total observed radiation $L_{tot} = L_{AGN} + L_{host}$. \cite{Shen:2011} provided a relationship between $L_{AGN}$ and $L_{host}$ which is suitable for the case of luminous QSOs
\begin{equation*}
    \frac{L_{5100,\,host}}{L_{5100,\,AGN}} = 0.8052 -1.5502x+0.9121x^{2}-0.1577x^{3}
\end{equation*}
where $x+44 \equiv \log(L_{5100,\,tot}/[erg\,s^{-1}])$. This formula is valid for $\log(L_{5100,\,tot}) < 45.053$. At higher luminosities the host contribution is supposed to be negligible. \\
For the WISSH sample we extrapolated  $L_{5100}$ by linearly interpolating the data from the two nearest bandpasses. All 85 sources have $\log(L_{5100}) > 45.053 $ (see fig. \ref{fig:mean_SED} and table \ref{tab:lbol}) so no corrections for the host were applied.

\subsection{Interpolation and construction of the mean SED}
\label{sec:interpolation}

We linearly interpolated (in the \textit{log-log} space) the individual SEDs in order to have them binned equally.
In detail,  we interpolated the luminosity points over a grid  of width $\Delta (\log \lambda) = 0.02$. The errors on the interpolated points were computed linking the upper (or lower) bounds of the adjacent observed luminosities.
The mean SED for each wavelength of the grid $\overline{\lambda L}$ was subsequently derived as the weighted geometric mean of the points
\begin{equation*}
    \overline{\lambda L} = \exp\left(\frac{\sum_{i}^{N}\log(\lambda L_{i})w_{i}}{\sum_{i}^{N}w_{i}}\right)
\end{equation*}
where $w_{i} \equiv \left(\lambda L_{i}/\sigma_{i}\right)^{2}$ are the weights and $\lambda L_{i}$ and $\sigma_{i}$ are the interpolated luminosities and their associated errors respectively.\\
Errors on the average SED have been evaluated as the geometric variance
\begin{equation*}
    \sigma^2 = \frac{\sum_{i}^{N}\log(\frac{\lambda L_{i}}{\overline{\lambda L}})^2}{N-1}
\end{equation*}
and the confidence interval is given as
\begin{equation*}
    \lambda L_{max}, \,\lambda L_{min} =\exp\left(\overline{\lambda L} \pm \frac{\sigma}{\sqrt{N}}\right)
\end{equation*}
Since 4 of the 16 sources with available  \textit{Herschel} photometry have at least one band where they are undetected (usually the 500 \textmu m one, where only upper limits on their fluxes are available), we performed \textit{gap repair} using our own SED; in detail we first built a mean SED using only the remaining  12 QSOs with complete \textit{Herschel} detections and used this one to replace the upper limits with a fixed luminosity value.
The \textit{gap repair} SED was also used for the 4 QSOs with available FIR \textit{ALMA} or \textit{NOEMA}  data but lacking \textit{Herschel} observations and to reconstruct the FIR emission of the 3 sources undetected by \textit{NOEMA} and for which we only have upper limits.
Finally we derived rest-frame 10 \textmu m  repaired photometry for all 85 QSOs with the aim of avoiding abrupt changes in the mean SED shape due to the gradually decreasing number of sources due to their different redshift distribution. It should be emphasized that this last \textit{photometry repair} is only intended to provide a 'smooth' SED template in the range 9 \textmu m $\lesssim \lambda \lesssim$ 50 \textmu m where, in the absence of observational data, we are forced to reconstruct the shape of the SED. For this wavelength interval, our mean SED cannot be considered as representative of the true emission by highly luminous QSOs. This also explains why the prominent $\sim 10$ \textmu m Silicate feature commonly observed in QSOs spectra \citep[e.g.][]{Siebenmorgen:2005} is not visible in our mean SED.
\subsection{X-ray and EUV range}
\label{sec:x_and_EUV}
\begin{figure}
    \centering
    \includegraphics[width= 0.4\textwidth]{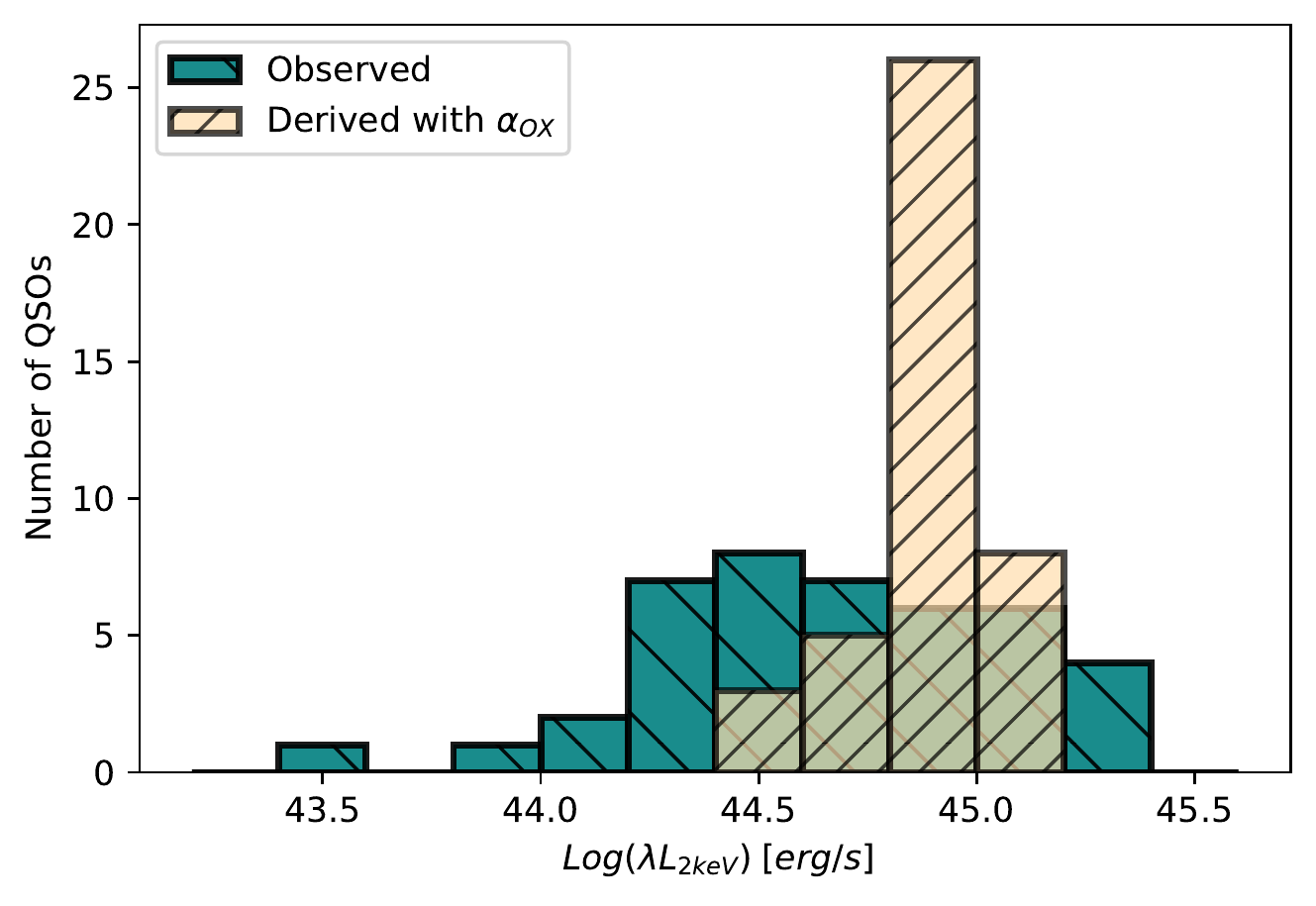}
    \caption{2 keV luminosity distribution of the sample. We include both measured values (green histogram) by \cite{Martocchia:2017} and \cite{Zappacosta:2020} and derived values using the $L_{2keV}-L_{2500\text{\AA}}$ relation.}
    \label{fig:Xdata_hist}
\end{figure}
\begin{figure*}
    \centering
    \captionsetup{width=0.8\linewidth}
    \includegraphics[width=0.8\textwidth]{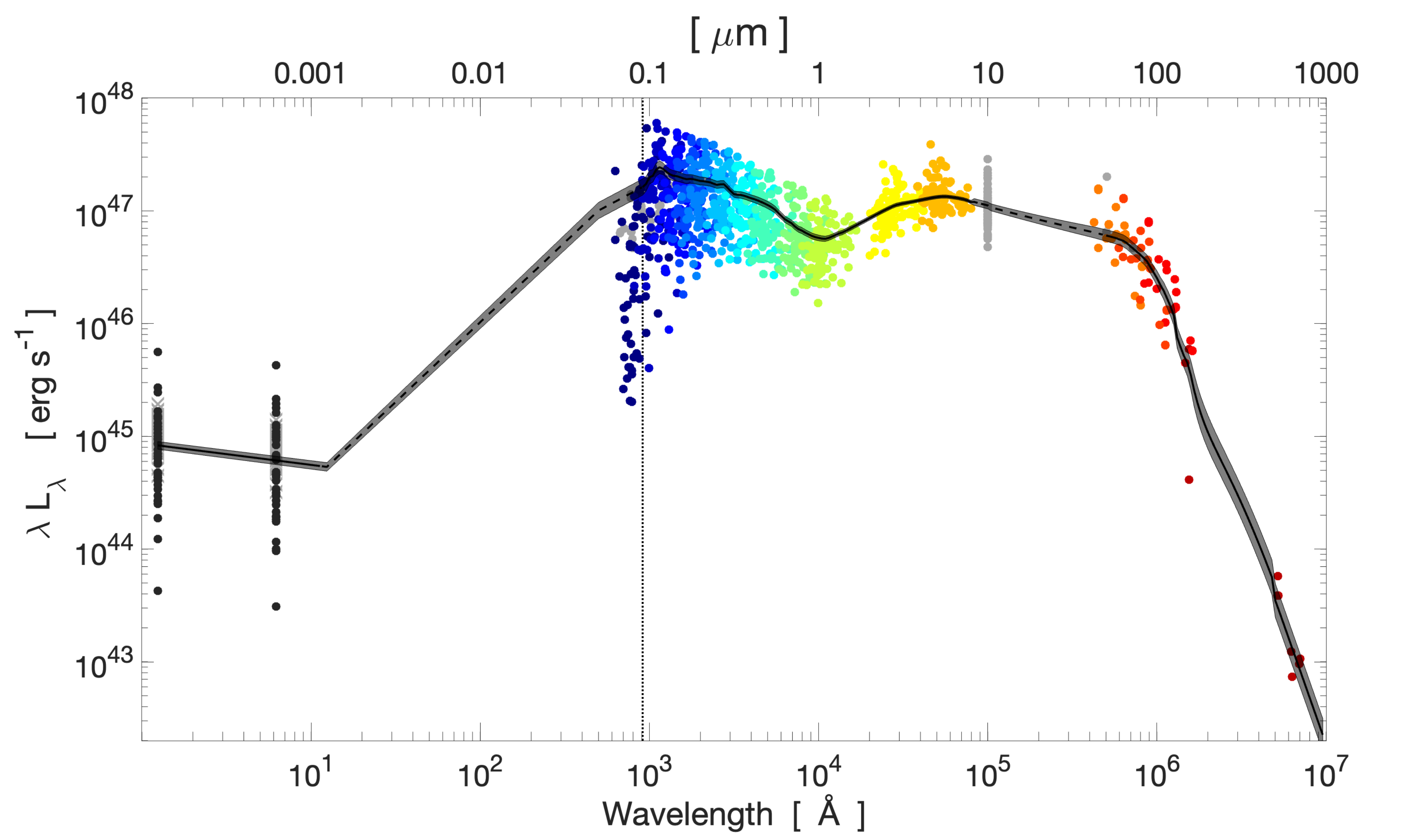}
    \caption{Mean SED derived from the WISSH sample. The shaded area gives the 68\% confidence interval. The colored circles represent available data points, color-coded according to the filter in which they were observed. Photometry obtained by \textit{Gap Repair} is represented as grey circles. The grey crosses indicate X-ray luminosities reconstructed from the $\alpha_{OX}$ relationship.
    The dotted black line indicates the Lyman limit $\lambda_{L} =912\,\text{\AA}$: for shorter wavelengths the mean SED was truncated and extended as explained in Sec. \ref{sec:x_and_EUV}.}
    \label{fig:mean_SED}
\end{figure*}

To characterize the mean SED in the X-ray region we employed 
 photon indexes $\Gamma$ recovered in the work by \cite{Martocchia:2017} and \cite{Zappacosta:2020} to transform the \textit{absorption corrected} integrated luminosities into 2 and 10 keV monochromatic luminosities, assuming a power law trend $L_{\nu} \propto \nu^{1-\Gamma}$. For sources for which we do not have X-ray data we used the known relationship between $L_{2500\text{\AA}}$ and the $\alpha_{OX}$ 
 \citep[][]{Vignali:2003,Steffen:2006, Lusso:2010} to estimate the 2 keV luminosity. In detail, we employed the relationship provided by \cite{Martocchia:2017}, which is in excellent agreement with the others  in the literature and has been derived using also the WISSH sample, and derived $L_{2500\text{\AA}}$ by linearly interpolating two adjacent luminosity points. For these sources  we then estimated the 10 keV luminosity assuming a standard value $\Gamma = 1.8$ \citep[][]{Piconcelli:2005,Martocchia:2017}. Given the wide spread of the observed X-ray luminosities, including these reconstructed values does not affect the mean SED (see fig. \ref{fig:Xdata_hist} and also fig. \ref{fig:mean_SED} where reconstructed X-ray luminosities are plotted as grey crosses).\\
 Between the X-ray band and the bluer UV filter, there is a rather wide unsampled region of the spectrum; in fact, for $100\,\text{\AA} \lesssim \lambda \lesssim 1200\,\text{\AA}$ , and especially below the Lyman limit ($\lambda_{L} = 912\, \text{\AA}$), the absorption by hydrogen clouds is responsible for the almost total absorption  of the emitted photons. These wavelengths are around the peak of AGN emission (the so-called Blue-Bump from the accretion disk).
Since computing the bolometric luminosities of the QSOs is among the goals  of this work and $L_{bol}$ is given by the integral of the SED, it is necessary to find a way to extrapolate the shape of the latter in this  extreme UV region.\\
In the literature there are several works dedicated to the characterization of the AGN SED in the EUV  \citep[][]{Mathews:1987, Korista:1997, Scott:2004, Casebeer:2006, Stevans:2014, Lusso:2015}; we opted for the adoption of the broken power-law discussed by \cite{Lusso:2010} as it uses observations by \cite{Zheng:1997} to define the QSO average spectrum up to $\lambda =500\;\text{\AA}$ which, despite not being among the most recent ones, provides a more conservative estimate of the flux (i.e. it has a steeper slope). In detail, we truncated the mean SED at the Lyman limit and proceeded to extend it with a fixed slope power-law  with $\alpha_{\lambda} = 0.8$ up to $\lambda = 500\,\text{\AA}$; we then linearly connected the end of this curve with that of the 1-10 keV power-law. Thus, the X-ray to UV region of the mean SED takes the form:

\begin{equation*}
\label{eqn:lusso_extension}
\lambda L_{\lambda}\propto\begin{cases}
\lambda^{-0.206}\;\text{at}\; 1.24\,{\text{\AA}} \leq \lambda \leq 12.40\,{\text{\AA}}\\
\lambda^{1.168}\;\text{at}\;  12.40\,{\text{\AA}}<\lambda \leq 500\,{\text{\AA}}\\
\lambda^{0.8}\; \text{at}\; \;500\,{\text{\AA}}<\lambda \leq 912\,{\text{\AA}}\\
\end{cases}
\end{equation*}
The characterization of the EUV band directly affects the measured value of $L_{bol}$. Since several other authors \citep[including K13 and][with whom we compare]{Runnoe:2012} prefer to use a single power law to link the UV to the X-rays, we quantified the impact of using a double PL rather than a single spectral slope, finding that in our case the former option leads to a 5\% higher $L_{bol}$ measurement.

The resulting full mean SED from the WISSH sample, extending from $10$ keV to $\sim 1000$ \textmu m, is shown in fig. \ref{fig:mean_SED} and given in table \ref{tab:SED}.  
We indicate with a dashed line those parts of the mean SED that were reconstructed.
\section{Results}

\subsection{Comparison with previously derived AGN mean SED}
\label{sec:comparison}
In this section we compare the shape of the derived mean SED with what reported in literature. Given the extreme luminosities of the WISSH sample compared to the bulk of the population, to ease the comparison, it is 
necessary to normalize the SEDs at a given wavelength. In this respect, the comparison has to be intended as 
relative differences, obtained assuming an equal luminosity at the normalization wavelength.
\begin{figure*}
    \centering
    \captionsetup{width=.8\linewidth}
    \includegraphics[width=0.8\textwidth]{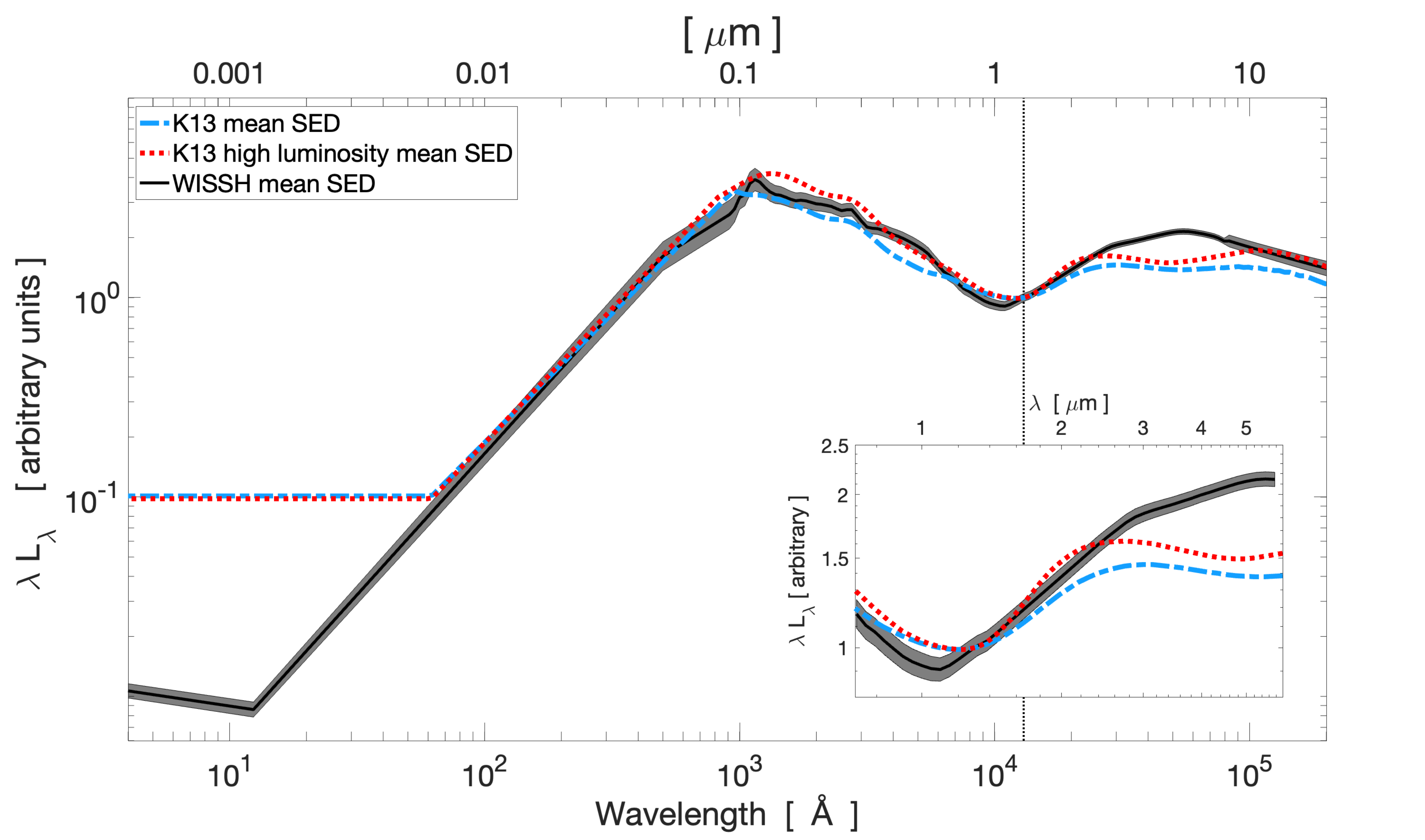}
    \caption{Comparison between the WISSH mean SED derived in this work (\textit{black line}), the overall mean SED for type 1 QSOs by K13 (\textit{light blue line}) and the high-luminosity SED also by K13 (dotted red line). SEDs are normalized at $1.3$ \textmu m (dotted vertical line). The box shows the zooming  of the SEDs in the $0.7<\lambda /\mu m<6$ range.}
    \label{fig:SED_comparison}
\end{figure*}
In fig.  \ref{fig:SED_comparison}, we show the comparison of the WISSH SED with the overall mean SED by K13, which, in addition to being one of the statistically most robust composite SEDs in the literature, was derived following a procedure very similar to the one adopted here. Therefore the SED by K13  is particularly suitable for making a comparison  between  the typical emission of  type 1 QSOs and that by our hyper-luminous sources. We  also include in the plot the mean SED derived by K13 for their high luminosity  subsample ($\log(\lambda L_{2500\text{\AA}}/[erg\,s^{-1}]) \geq 45.85$, corresponding to $46\lesssim\log(L_{bol}/[erg\,s^{-1}])\lesssim47.5$). All SEDs are normalized at $\lambda = 1.3$ \textmu m which corresponds approximately to the dip between the accretion disc UV bump and the obscuring torus MIR emission. To make a comparison which is independent of the wavelength chosen for normalization, in our analysis, where possible, the comparison was made either in terms of spectral slopes (i.e., luminosity ratios) or in terms of the percentage of emission of one component relative to another. \\
Overall, SEDs are quite similar in shape and, as expected, differences are marginally less pronounced when the comparison is made with respect to the high luminosity mean SED by K13. There are, however, some significant discrepancies, in particular in the X-ray and in the near and mid-IR bands. 
\subsection{X-ray to Optical bands}
At high energies, we find that the SED of hyper-luminous quasars has a lower X-ray emission compared to that of the bulk of the AGN population. This is due to the nonlinear relationship between the UV and X-ray luminosity, expressed through the $\alpha_{OX}$, and more precisely to the well known  anti-correlation between $\alpha_{OX}$ and $L_{2500 \text{\AA}}$ (see references in Sec. \ref{sec:removing_red}) meaning that UV luminous QSOs are \textit{relatively} X-ray weak. Indeed we find that $\alpha_{OX} = -1.94$ for the WISSH mean SED ($\alpha_{OX} = -1.91$ considering only sources with X-ray observed data), while for comparison, the K13 mean SED  has   $\alpha_{OX} = -1.53$. It should be emphasized that these values, being $\alpha_{OX}\equiv \frac{\log (L_{2keV}/L_{2500\text{\AA}})}{2.605}$, do not depend on the parametrization chosen to reconstruct the EUV SED.\\
Smaller differences (within the $3\sigma$ confidence interval, assuming the same $L_{1.3\mu m}$)  between the WISSH mean SED and the K13 one have been found also in the UV region, between the Ly$\alpha$ and Lyman limit, where we observe a decline in the emission of hyper-luminous quasars that is not present in neither of the two SEDs provided by K13. This can likely be attributed to our conservative estimates made in terms of IGM absorption (i.e. not considering Damped Ly$\alpha$ systems which could instead be relevant especially for our high-z QSOs) and to a redshift effect on the edge. 
Also, in the optical band, approximately between $\lambda =3000\,\text{\AA}$ and $8000\,\text{\AA}$, we note that both WISSH mean SED and K13 high luminosity SED have a steeper continuum. Indeed, WISSH mean SED and K13 high-luminosity subsample have $\alpha_{opt} = 0.8\pm0.1$ and 0.85, respectively, to be compared with $\alpha_{opt} = 0.65$ exhibited by the whole K13 sample. Such a difference could be addressed to a primary UV component having higher temperatures, as could be expected in this class of highly luminous AGN.

\subsection{IR region}
In the near and mid infrared, for $\lambda > 1$ \textmu m, the WISSH mean SED shows 1) a more prominent bump at 3-9 \textmu m  with a $\sim5\sigma$ significance at $5$ \textmu m (again, assuming the same $L_{1.3\mu m}$), and 2) a dip 
shifted to slightly shorter wavelengths, i.e. $\lambda_{dip} \approx 1.1$ \textmu m, while in both SEDs  by K13 \citep[and also those by][]{Elvis:1994,Richards:2006} it is found at  $\lambda \approx 1.3$ \textmu m. Although providing a  statistical significance to this shift in wavelength is not easy, we are confident that it is not an artifact nor it depends on the method adopted, since it is 4 times the resolution of the grid adopted to interpolate the data points and the procedures adopted by \cite{Richards:2006} or K13 to derive their mean SEDs are fairly analogous to ours. However, there are at least two factors that we cannot rule out that may have influenced our result; firstly, the limited number of WISSH QSOs may not be sufficient to adequately sample a region where a sudden change in slope occurs. Moreover, our results could be influenced, at least partially, by host emission which peaks around those wavelengths and that we assumed is negligible (see sec. \ref{sec:host_correction}). The change of the dip position will be the focus of a future work (\textit{Saccheo et al. in prep.}).\\ \\
We quantified the WISSH NIR to MIR excess by using the ratio between the integrated IR and optical emission, i.e. $R\equiv L_{1.3-9 \mu m}/L_{1216\text{\AA} -1.3\mu m}$. We found that, for WISSH QSOs, $R$ is $\approx 32 \%$ higher than that obtained from K13 high-luminosity SED.
Alternatively, by normalizing the IR emission to the luminosity at the dip between the disk and torus bumps, (i.e. $L_{1.3-9 \mu m}/L_{dip}$) we get a $\sim 29 \%$ excess, which is consistent with the previous result.
The stronger bump is not unexpected since the sources of the sample were specifically selected  to be  the most luminous at $\lambda = 7.8$ \textmu m.  
Moreover both works by \cite{Richards:2006} and K13 show that luminous quasars tend to have a relatively stronger IR emission. Also \cite{Duras:2017}, analyzing a sub sample of 16 WISSH quasars, had already pointed out that in about 30\% of the sources an additional emission component  was needed to properly model the near to mid IR emission; they accounted for this IR excess with an extra hot dust component with temperature ranging from 650 to 850 K depending on the source. 
However, in none of these works a shift of the dip was reported.

\begin{figure}
    \centering
    \includegraphics[width = 0.4\textwidth]{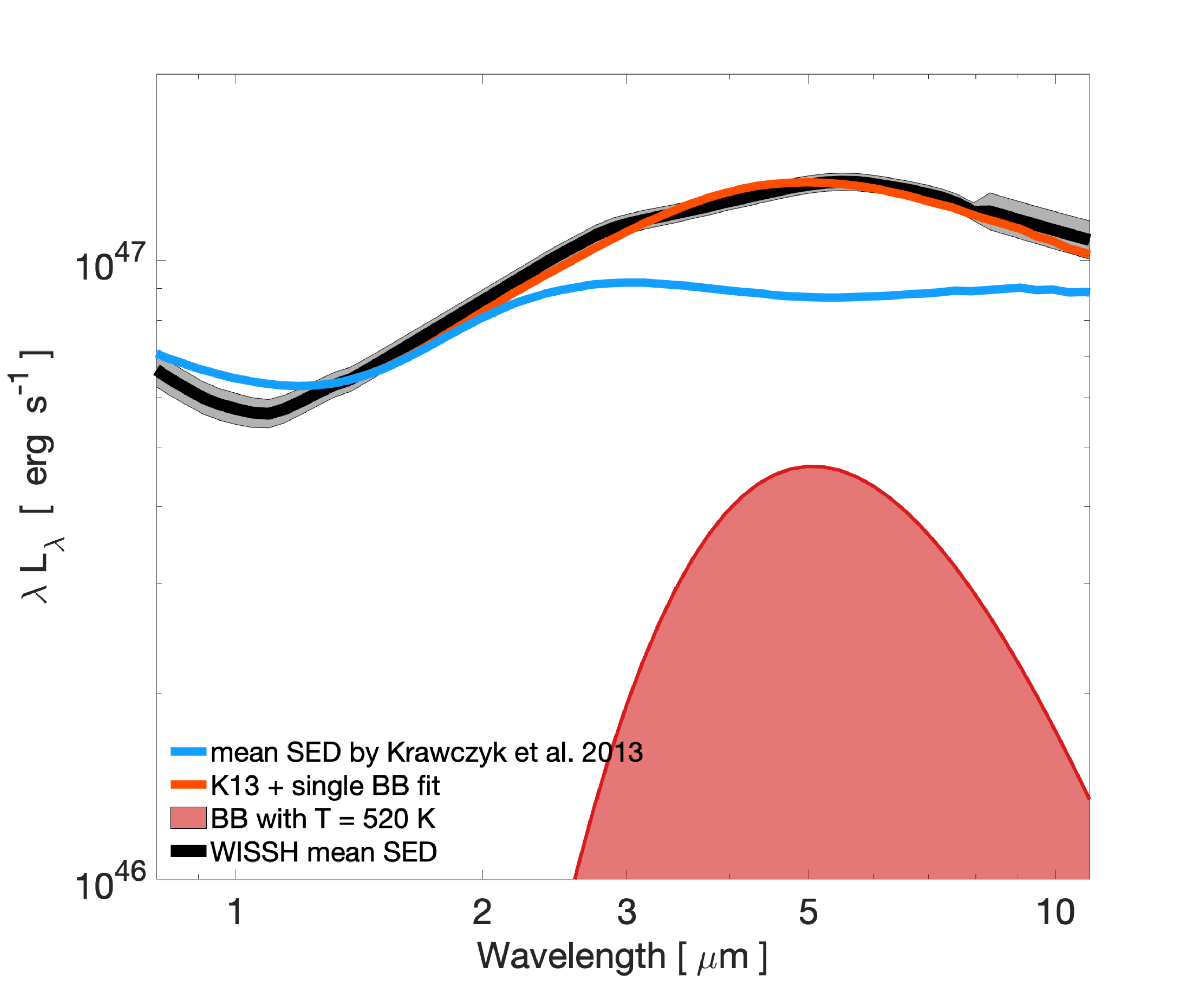}
    \includegraphics[width = 0.4\textwidth]{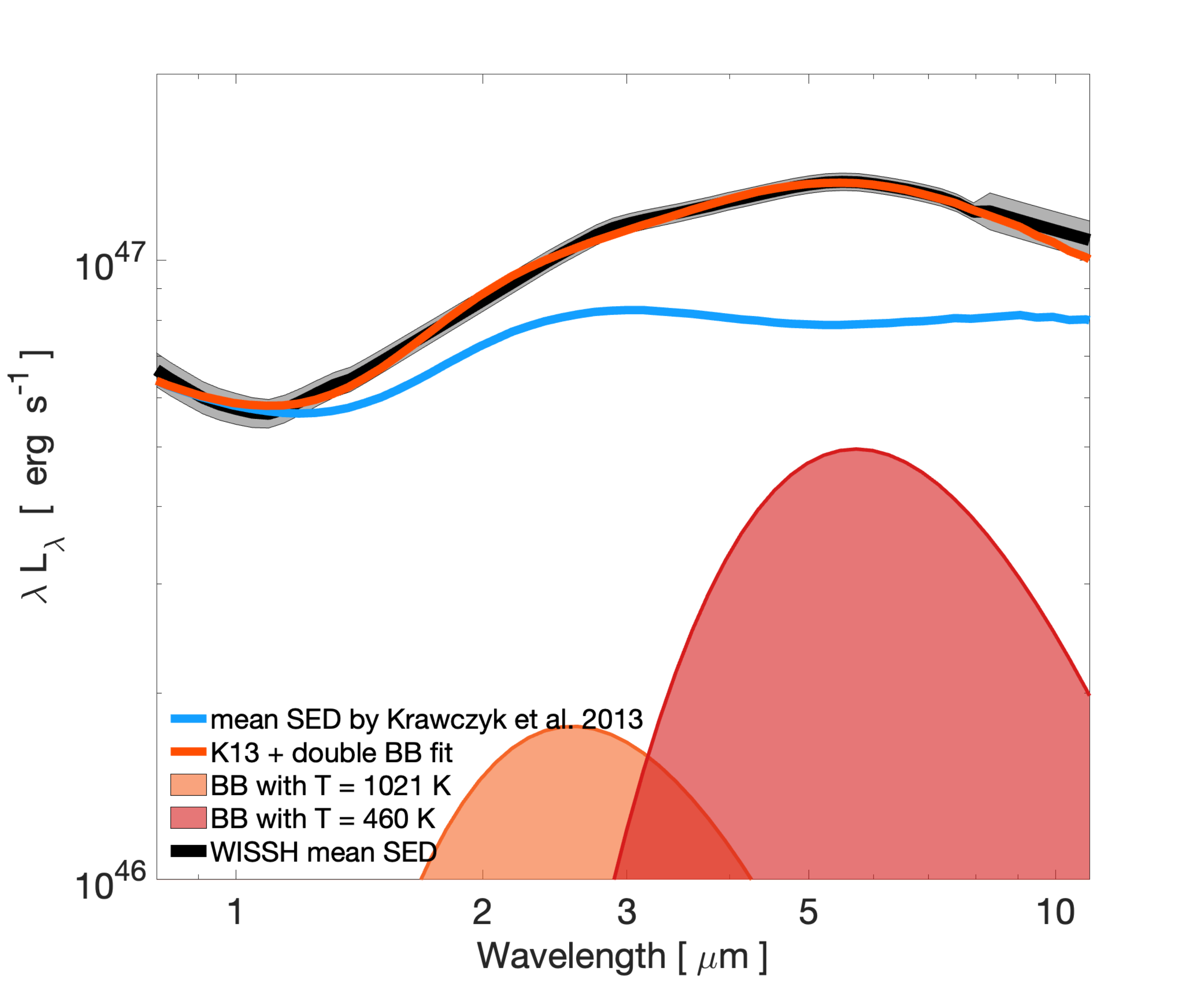}
    \caption{Modeling of the WISSH mean SED via the K13 composite SED plus one (top panel) or two (bottom panel) modified black-bodies as described in Sec. \ref{sec:comparison}. Comparing the two solid orange lines it is evident how the addition of a hotter dust component allows a better reproduction of the curve for $\lambda <2$ \textmu m.}
    \label{fig:black_bodies}
\end{figure}
To further verify the hypothesis of an extra hot dust component, we tried to reproduce the WISSH SED curve at $0.8 <\lambda/\mu m < 8$ by fitting a combination of the mean SED by K13 plus a modified black body with temperature free to vary in the range 150-1600 K as follows: 
\begin{equation*}
    F_{\nu}(\lambda)^{WISSH} = aF_{\nu}(\lambda)^{K13} +bB_{\nu}(\lambda, T)
\end{equation*}
where $a$ and $b$ are the relative normalizations.
The best fit, according to $\chi^{2}$ minimization, is obtained with the addition of a BB at $T=520$ K and is shown as the solid orange line in the top panel of fig. \ref{fig:black_bodies}. As visible, this extra dust component optimally describes the mid IR bump but fails to explain the shifting of the dip, since the emission of a black body with such a temperature is negligible at those wavelengths: a hotter component is therefore required.  
As a next step, we shape the mean SED by including two black bodies with two distinct ranges of possible temperatures, $T_{hot} \sim 900-1600$ K, and $T_{warm} \sim 150-900$ K \citep{Hernan-Caballero:2016}, i.e.:
\begin{equation*}
    F_{\nu}(\lambda)^{WISSH} = aF_{\nu}(\lambda)^{K13} +bB_{\nu}(\lambda, T_{hot}) +cB_{\nu}(\lambda, T_{warm})
\end{equation*}
The best fit solution, shown in  the lower panel of fig. \ref{fig:black_bodies}, includes a BB at $T_{hot} = 1021$ K and one at $T_{warm} = 460$ K\footnote{These values should be intended as purely indicative as in reality we expect a gradient of temperatures.}. Both the NIR excess and the shift of the dip are now reproduced. The WISSH sources have therefore an IR excess with respect to the mean QSOs SED which is due to two distinct contributions: 1) a higher emission by a  warm dust component ($T\sim 450-800$ K) heated by the AGN and likely associated to the obscuring torus which is explainable as larger covering factor of the torus itself; 2) an extra emission by very hot dust ($T>1000$ K) which is responsible for the NIR excess and the shifting of the dip.\\
In the attempt to provide a physical explanation and account for these extra hot and warm dust components, we used SKIRTOR by \cite{Stalevski:2012} \citep[in the version implemented in CIGALE,][]{Yang:2022} to generate several AGN SED templates and fit the mean SED between $\lambda=1216\,\text{\AA}$ and $\lambda = 8$ \textmu m. In detail, we narrowed the parameter space by fixing  the viewing angle to 0, since the strong NIR emission suggests that the inner layer of the torus is directly visible \citep[see also][]{Duras:2017}. 
\begin{figure}
    \centering
    \includegraphics[width=0.5\textwidth]{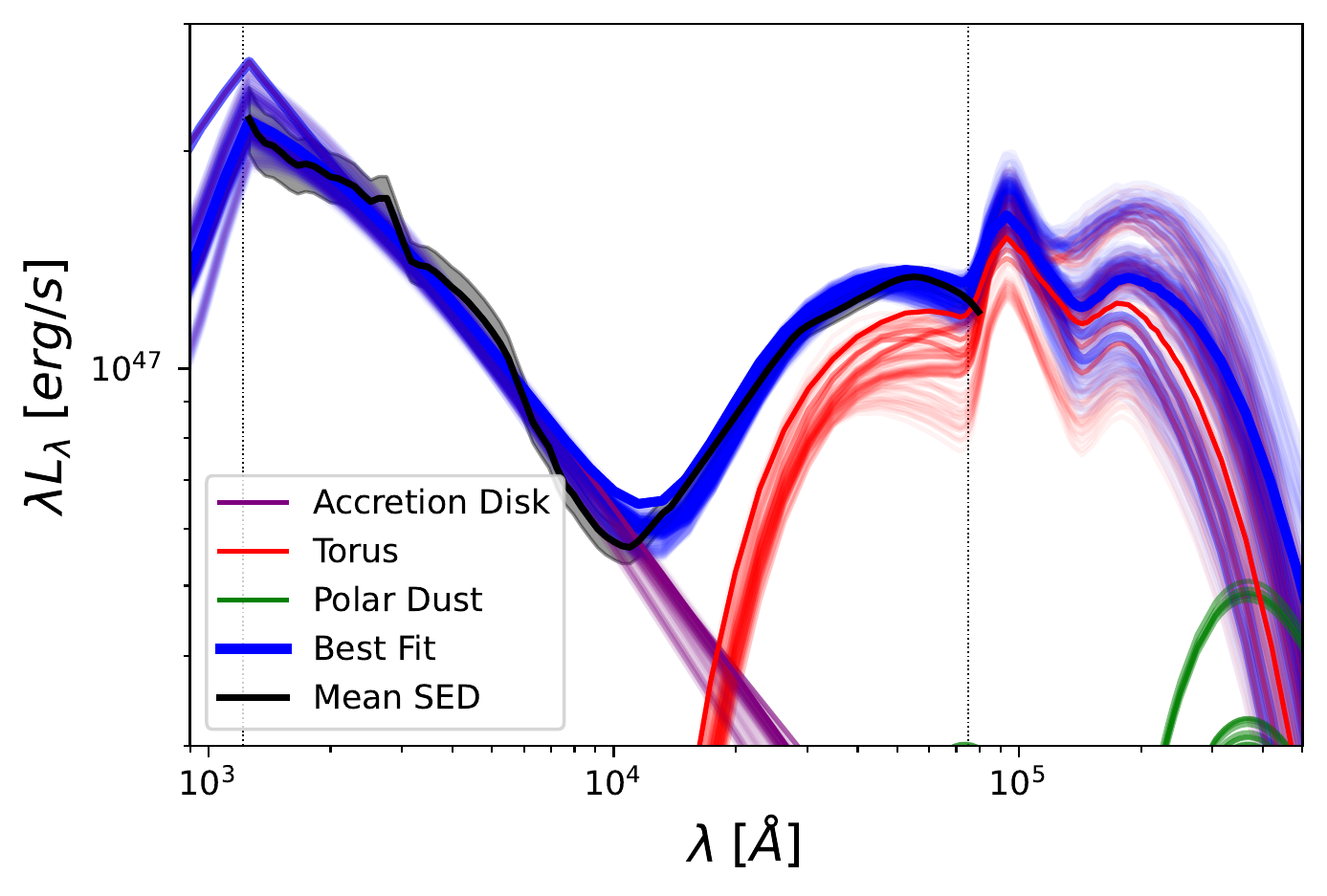}
    \caption{Best-fit and 1-$\sigma$ errors of the WISSH mean SED via CIGALE-generated templates. For illustrative purposes the different components considered in CIGALE and whose sum provides the overall AGN SED are plotted with different colors. The best-fit parameters are: $OA = 50^{\circ}$, $\tau_{\nu}$=11, p=1.0, q=0, $\delta_{AD}$=-0.4, $R_{in}/R_{out}=20$, $T_{PD}$=100 K, $E_{B-V} =0.03$; we refer to \cite{Yang:2022} for a detailed description of these parameters.
    The dotted lines delimit the wavelength range over which the fitting was performed.}
    \label{fig:cigale_fit}
\end{figure}

The best-fit results are reported in fig. \ref{fig:cigale_fit} and show that, although there is very good agreement at $\lambda\gtrsim 1.5$ \textmu m, this is not true for wavelengths around the minimum where no model seems to be able to describe the shape of our SED. 
We have also considered a polar dust component since in some cases it has been found to be necessary to explain the strong NIR emission of HDO AGN \citep[e.g.][]{Lyu:2018}. However, a large contribution from this component would be associated with a strong extinction of the optical continuum that we do not observe in our QSOs; indeed, in our fit the polar dust gives a completely negligible contribution and therefore cannot represent the extra hot dust component discussed above. Therefore, while the extra emission that we attributed to an additional warm dust component in our empirically-driven reasoning can be justified as a specific configuration of the torus geometry, we are unable to explain the nature of the hot component.


\subsection{Bolometric Luminosities and Bolometric corrections}
\subsubsection{Bolometric Luminosities}
\label{sec:bolometric_lum}
The Bolometric luminosity is defined as the integrated area below the non-extincted SED and provides a measure of the energy budget of the AGN at all wavelengths (see below for details).\\
Fluxes, which are the actual measured physical quantity, are converted into luminosities under the hypothesis that the sources emit isotropically:
\begin{equation*}
    \nu L_{\nu}  = 4\pi D_{L}^{2} \nu F_{\nu}.
    \label{eqn:isotropic}
\end{equation*}
Under the isotropic assumption,  when computing $L_{bol}$, it is necessary to consider only the radiation emitted along the line of sight and remove from the calculation the reprocessed one, i.e. that of the obscuring torus, which is heated by photons originally emitted in another direction and whose contribution is already included in the $4\pi$ factor in the above equation. To avoid counting the same contribution twice, several authors \citep[e.g][]{Marconi:2004,Nemmen:2010, Lusso:2012,Runnoe:2012, Duras:2020} limit the integral under the SED to $\lambda < 1$ \textmu m. Regarding the high-energy integration limit, several authors \citep[e.g K13,][]{Duras:2020} make a similar reasoning as seen for the IR and place it in the soft X-ray band (between 0.5 and 2 keV) to avoid counting radiation reprocessed by the corona and consider only the proper accretion luminosity. However, in our case, to be consistent with \cite{Runnoe:2012} we decided to set the limit at 10 keV.\\
To derive $L_{bol}$ we model  each source emission data points with the derived mean SED plus a dust extinction component \citep[see e.g.][]{Bongiorno:2012}:
\begin{equation*}
    F_{\lambda}^{QSO}(\lambda) = aF_{\lambda}^{model}(\lambda)\times 10^{-0.4A(\lambda )}
\end{equation*}
where the extinction $A(\lambda )$ is computed assuming the SMC dust reddening law by \cite{Prevot:1984} (i.e. $A(\lambda ) = 1.39\lambda^{-1.2}E_{B-V}$, $\lambda$ in \textmu m). By $\chi^2$ minimization procedure with respect to the observed luminosity points in the range $1216\,\text{\AA} \leq\lambda \leq 10$ \textmu m we obtain the best values for the normalization $a$ and the color excess $E_{B-V}$. We set the lower limit to $1216\,\text{\AA}$ because for shorter wavelengths the drop in the observed flux is not  completely attributable to dust extinction but also to IGM absorption. $L_{bol}$ is then derived by integrating the normalized non-extincted mean SED between 1 \textmu m and 10 keV. \\
To calculate the associated errors first, if needed, we gradually increased the uncertainties on the luminosity points  until we met the condition $\chi^{2}_{best} \sim 1$ (e.g. \citealt{Gruppioni:2008}, but see also \citealt{Andrae:2010} for a discussion about the underlying assumptions and the shortcomings of this procedure).
Then we found the lower and the upper errors as the minimum and maximum values among the models that satisfy the condition $\chi^{2}- \chi^{2}_{best} <1$; to these uncertainties  we add in quadrature  those related to the $L_{bol}$ of the mean SED, calculated as the difference obtained if we integrate the lower (or upper) bounds instead of its average value. We also consider an additional contribution to the overall uncertainties, given by the difference between $L_{bol}$ computed using the mean SED and that derived  by connecting their intrinsic luminosity point with straight lines. This component is particularly relevant for QSOs with $E_{B-V} =0$ and which have an intrinsic bluer SED than the average one; this way we take into account that their  $L_{bol}$ might be underestimated.

\begin{table*}
\begin{center}
\begin{tabular}{cccccc}

ID & $Log(L_{bol})$ & $Log(L_{2500\text{\AA}})$ & $Log(L_{5100\text{\AA}})$ &        $Log(L_{3\mu m})$ & $E_{B-V}$\\
\hline
\hline
WISSH01 &  $47.53_{-0.13}^{+0.08}$ &  $47.08 \pm 0.01$ &    $46.77 \pm 0.03$ &    $46.94 \pm 0.01$ &   0.06 \\
WISSH02 &  $47.55_{-0.05}^{+0.06}$ &  $47.02 \pm 0.06$ &    $46.82 \pm 0.07$ &    $46.9 \pm 0.05$ &    0.0 \\
WISSH03 &  $47.48_{-0.08}^{+0.07}$ &  $46.94 \pm 0.04$ &    $46.78 \pm 0.07$ &    $46.93 \pm 0.03$ &   0.02 \\
\hline
\end{tabular}
\end{center}
\caption{Bolometric and monochromatic luminosities for WISSH quasars. Units are in [$erg\,s^{-1}$]. Monochromatic luminosities are computed with a linear interpolation in the \textit{log-log} space and their uncertainties are computed as in sec. \ref{sec:interpolation}. Monochromatic luminosities are also intrinsic, i.e. they are corrected for dust extinction.
\\The full table is available in the online version.}
\label{tab:lbol}
\end{table*}

\begin{figure*}
    \centering
    \includegraphics[width=1\textwidth]{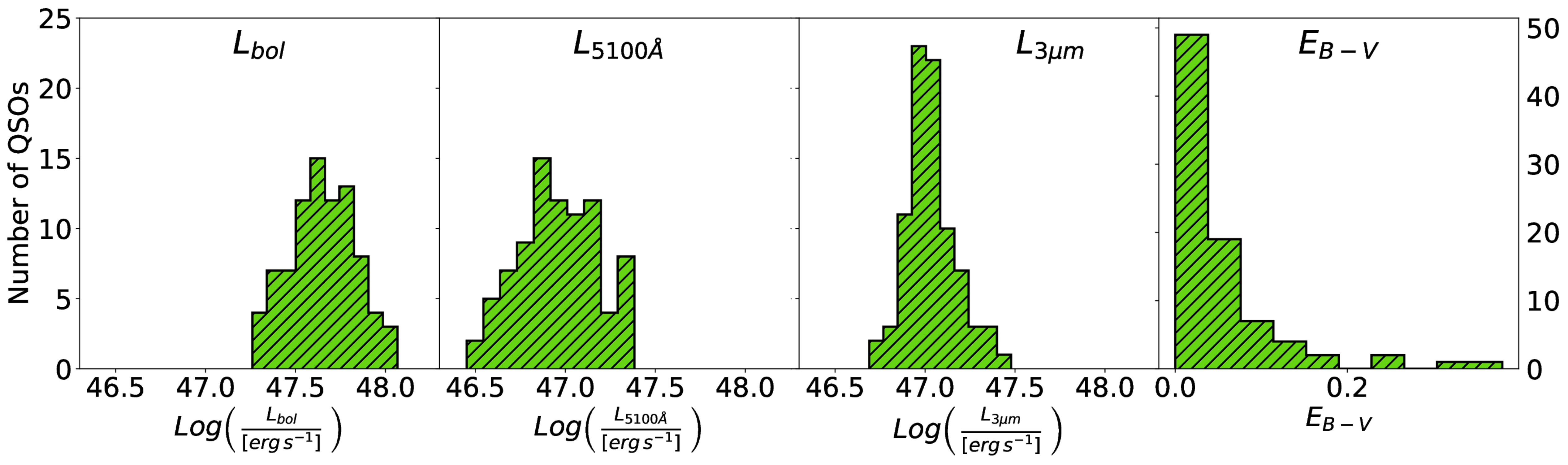}
     \caption{From \textit{left} to \textit{right}: Bolometric luminosities, intrinsic monochromatic luminosities at $5100\,\text{\AA}$  and $3\,\mu m$, and $E_{B-V}$}
     \label{fig:luminosities}
    
\end{figure*}

The derived $L_{bol}$ are reported along with intrinsic monochromatic luminosities at different wavelengths and the $E_{B-V}$ in table \ref{tab:lbol} and their distributions are shown in fig. \ref{fig:luminosities}.\\

\subsubsection{Bolometric corrections}
\label{sec:bolometric_corrections}
We then used the derived $L_{bol}$ to compute the bolometric correction, i.e. the ratio between the bolometric luminosity and the monochromatic luminosity in a specific band:  $K(\lambda) = \frac{L_{bol}}{\lambda L_{\lambda}}$. In particular we derived bolometric corrections for $\lambda = 5100\, \text{\AA}$ and 3 \textmu m, two bands where the WISSH mean SED shows differences with respect to the K13 SED. The purpose of our analysis is indeed twofold: on the one hand, we are interested in inspecting how WISSH QSOs compare with the bulk of the population and with other sources with comparable luminosity, on the other hand, we have the opportunity to study bolometric corrections in a luminosity range poorly explored so far. Indeed it is widely accepted \citep[e.g][]{Lusso:2012, Runnoe:2012, Duras:2020} that, in most bands, bolometric corrections are not constant but rather functions of $L_{bol}$. WISSH quasars are therefore excellent sources to investigate these dependencies at the bright end of the AGN luminosity distribution.
Notably \cite{Duras:2020} already included WISSH QSOs in their study of X-ray (2-10 keV) and $4400\,\text{\AA}$ bolometric corrections across 7 orders of magnitude of quasar luminosity. 
\begin{figure*}
    \centering
    \captionsetup{width=.8\linewidth}
    \includegraphics[width=0.8\textwidth]{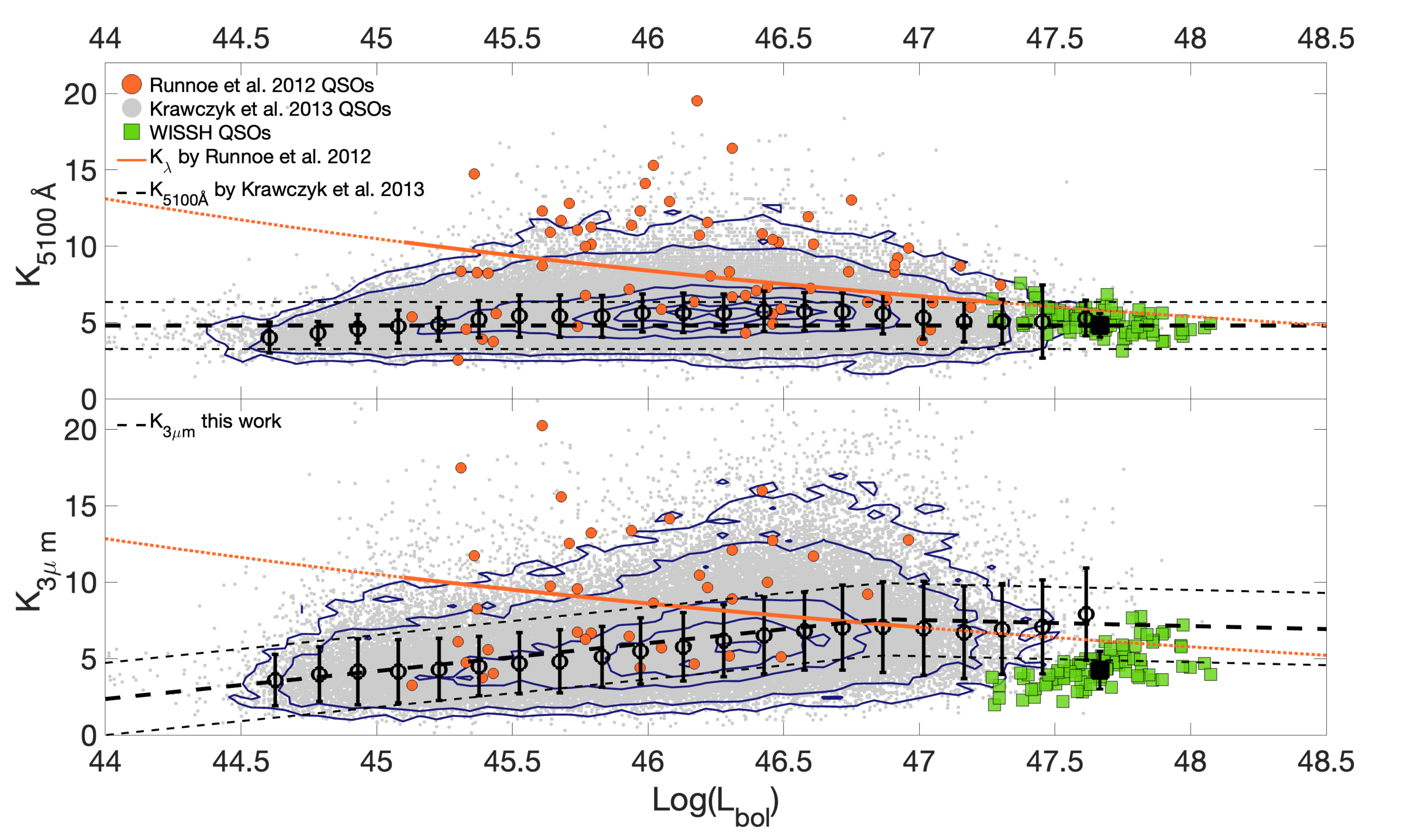}
    \caption{\textit{Top Panel:} $K_{5100\text{\AA}}$ vs $Log(L_{bol})$, units in [erg/s]. QSOs from \cite{Runnoe:2012} are marked as orange filled circles. Grey circles depict QSOs from \cite{Krawczyk:2013} while black circles provide their median values (with the associated spread) over bins of width $0.15\,dex$. WISSH QSOs are marked as green squares and the black square indicates their mean.
    The solid orange line describe the non-linear  relationship by \cite{Runnoe:2012}, the dotted line indicates that the relation is being extrapolated to lower or higher luminosities.
    The dashed black lines give the linear relationship by K13 and its associated uncertainty.\\ 
    \textit{Bottom Panel:} $K_{3\mu m}$ vs $Log(L_{bol})$, color-coded as above. Here the dashed black lines depict the best fit computed as explained in Sec. \ref{sec:bolometric_lum} and its uncertainty computed as the square root of the variance with respect to the best fit prediction.} 
    \label{fig:Bolometric_corrections}
\end{figure*}
In fig. \ref{fig:Bolometric_corrections} we show $K_{5100\text{\AA}}$ (\textit{upper panel}) and $K_{3\mu m}$ (\textit{lower panel}) vs $L_{bol}$; we have also included QSOs by \cite{Runnoe:2012} and K13. To give a clearer visualization of the position in the plot of the majority of QSOs by K13 we depict as black circles the median values obtained by grouping the sources in bins spaced $0.15\, dex$ in $L_{bol}$. It should be noted that by inserting all these sources in the same plot we are assuming as negligible any dependence on redshift.

In detail, \cite{Runnoe:2012}  derived bolometric corrections for a variety of wavelengths using 63 sources with $45 \lesssim \log\left(L_{bol}/[erg\,s^{-1}]\right) \lesssim 47$ at $ 0.3\leq z \leq1.4$. They found that non linear relationships in the form of $\log(L_{bol}) = A\log( L_{\lambda}) +B$ provided a better representation of the data \citep[see also][]{Nemmen:2010}. On the contrary, K13 reported a constant bolometric correction at $5100\,\text{\AA}$ without providing a value for $\lambda = 3$ \textmu m. 

In the case of $\lambda=5100\,\text{\AA}$, \cite{Runnoe:2012} found $\log (L_{bol}) = (0.91\pm 0.04)\log( L_{5100\text{\AA}}) +4.89(\pm1.66)$ (solid orange line) whereas K13 gives $K_{5100}=4.33\pm 1.29$\footnote{The $K_{5100\text{\AA}}$ reported by K13 was derived assuming 1 \textmu m and 2 keV as integration limits for $L_{bol}$. Including also the 2-10 keV luminosity we derived $K_{5100\text{\AA}}= 4.80 \pm 1.54$ which is the value shown in fig. \ref{fig:Bolometric_corrections} as a dot-dashed black line.}. As visible in the \textit{upper panel} of fig. \ref{fig:Bolometric_corrections} (solid orange vs dashed black lines), although these relationships are rather different in the low luminosity regime, they converge to comparable values for $L_{bol} \gtrsim 10^{46.5}$ erg/s and are both in agreement with the values derived from the analysis of the WISSH sample.
Indeed computing the mean bolometric correction on the WISSH sample gives $K_{5100\text{\AA}}= 4.8 \pm 0.8$ (shown as a black square in fig. \ref{fig:Bolometric_corrections}) which is consistent with both relations. 


In the case of $\lambda = 3$\textmu m, \cite{Runnoe:2012_IR} propose $\log(L_{bol}) = (0.92\pm 0.08)\log(L_{3\mu m}) +4.54(\pm 3.42)$ while, as anticipated, K13 does not provide a bolometric correction nor QSOs monochromatic luminosities. For these reasons, we have computed  $L_{3\mu m}$ of the K13 sample by linear interpolating the two closest bands. Data points, contours and median values reported in the \textit{bottom panel} of fig. \ref{fig:Bolometric_corrections} show a positive correlation between $K_{3 \mu m}$ and $L_{bol}$ up to $L_{bol}\approx 10^{46.7}-10^{46.9}$ erg/s followed by a flatter, or even decreasing, trend for higher luminosities. 
Therefore we fit QSOs by K13 with two linear functions joined at $L_{bol} = 10^{46.86}$ erg/s. We find $K_{3\mu m} = 1.83\log(L_{bol})-77.0$  when $L_{bol} \leq 10^{46.86}$ erg/s and $K_{3\mu m} = -0.39\log(L_{bol}) +26.04$ at higher luminosities. As before, while at low luminosities the K13 fit substantially differs from that by \cite{Runnoe:2012_IR} (the first increasing with $L_{bol}$, the other decreasing), at $L_{bol} >10^{47}$ erg/s they show a rather similar behaviour. 

An increasing  $K_{3 \mu m}$ ($\equiv \frac{L_{bol}}{L_{3\mu m}}$) with $L_{bol}$ is in agreement with \cite{Maiolino:2007} who found the $L_{IR}/L_{bol}$ ratio to be a decreasing function of $L_{bol}$ (more precisely they investigated $L_{5100\text{\AA}}$ and $L_{6.8\mu m}$ which, however, can be considered good tracers of $L_{bol}$ and $L_{IR}$ respectively). A similar result was also reported in \cite{Calderone:2012, Lusso:2013, Ma:2013} and \cite{Gu:2013}. In these papers the authors justify these results through the receding torus model \citep[e.g][]{Lawrence:1991, Simpson:2005} according to which the distance between the circumnuclear dust and the accretion disk increases with $L_{bol}$: as the QSOs luminosity increases, the covering factor decreases and, therefore, relatively lower emission is observed in the near and mid infrared bands. The observed high luminosity change of $K_{3\mu m}$ trend might then be interpreted as a limiting radius beyond which an increase in the primary emission does not correspond to a further receding of the torus but rather to an increment of dust heated close to its sublimation temperature. This is in agreement with what is found in the WISSH mean SED as well as by \cite{Gallagher:2007}, K13 and other authors who showed that luminous QSOs actually have a boosted IR emission.

Unlike what was observed for $K_{5100\text{\AA}}$,  for $\lambda =3$ \textmu m the  WISSH sample does not follow the general trend. Indeed, most of WISSH QSOs lie below the curves obtained by \cite{Runnoe:2012_IR} and that derived from K13 data. This result is attributable to the sample selection criterion that specifically collected the brightest sources in the mid-IR. 
We also note that for WISSH QSOs, $K_{3\mu m}$ increases with their $L_{bol}$. This is partially expected given the sample selection (the distribution of $L_{3\mu m}$ is narrower than $L_{bol}$, see fig. \ref{fig:luminosities}); however, a physical explanation for what we observe could be the fact that, although WISSH sources represent the lower envelope of the distribution, and thus are the QSOs with the highest contribution by hot dust, emission by this component is limited by the amount of available dust. It could be possible that, going to such extreme luminosities there is lack of additional dust to heat either to simply balance the increase in primary emission or to counteract the action of the receding torus. Hence the increasing trend in WISSH  $K_{3\mu m}$.
However, due to the very limited number of sources at those extreme luminosities, it is complicated to perform an analysis to validate this hypothesis on more solid statistical grounds; moreover it assumes that the luminosity-weighted estimated dust contribution is related to its mass distribution, which is not necessarily true.
\begin{figure}
\centering
\includegraphics[width =0.5\textwidth]{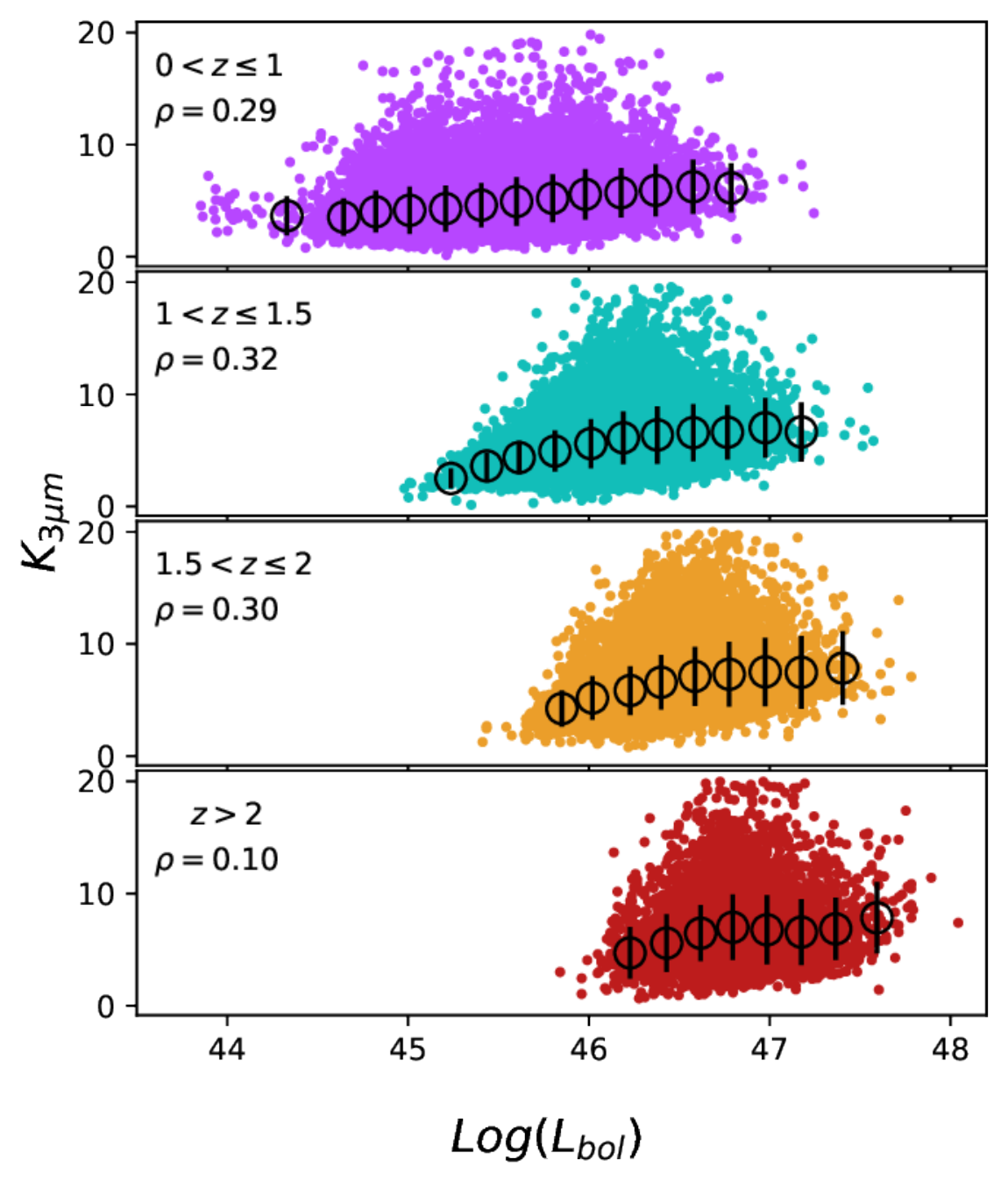}
\caption{$K_{3\mu m}$ for K13 QSOs binned according to their redshift. Black circles give their median values over bins of width $0.2\, dex$. Each panel shows the QSOs redshift range and the coefficient derived performing a Spearman correlation test.}
\label{fig:redshift_evolution}
\end{figure}

In fig. \ref{fig:redshift_evolution} we explore the possibility of a redshift dependence of $K_{3\mu m}$ by grouping K13 sources into 4 redshift bins ($z\leq1$, $1<z\leq1.5$, $1.5<z\leq2$ and z>2). The behavior of the bolometric correction seems to depend more on the luminosity rather than on the considered redshift range; in all bins the trend seems to be the same (increasing up to $L_{bol}\approx 10^{47}$ erg/s followed by a flatter trend) and the differences between the distributions can be attributed to the different number of available sources with a given luminosity in each redshift bin. This is also evident by looking at the results of a Spearman Correlation test: the first 3 bins have  similar correlation coefficients ($\rho\sim 0.3$) while the fourth, which contains many high luminous QSOs has one with a substantially lower value ($\rho = 0.10$).\\ 
\begin{figure}
    \centering
    \includegraphics[width=0.5\textwidth]{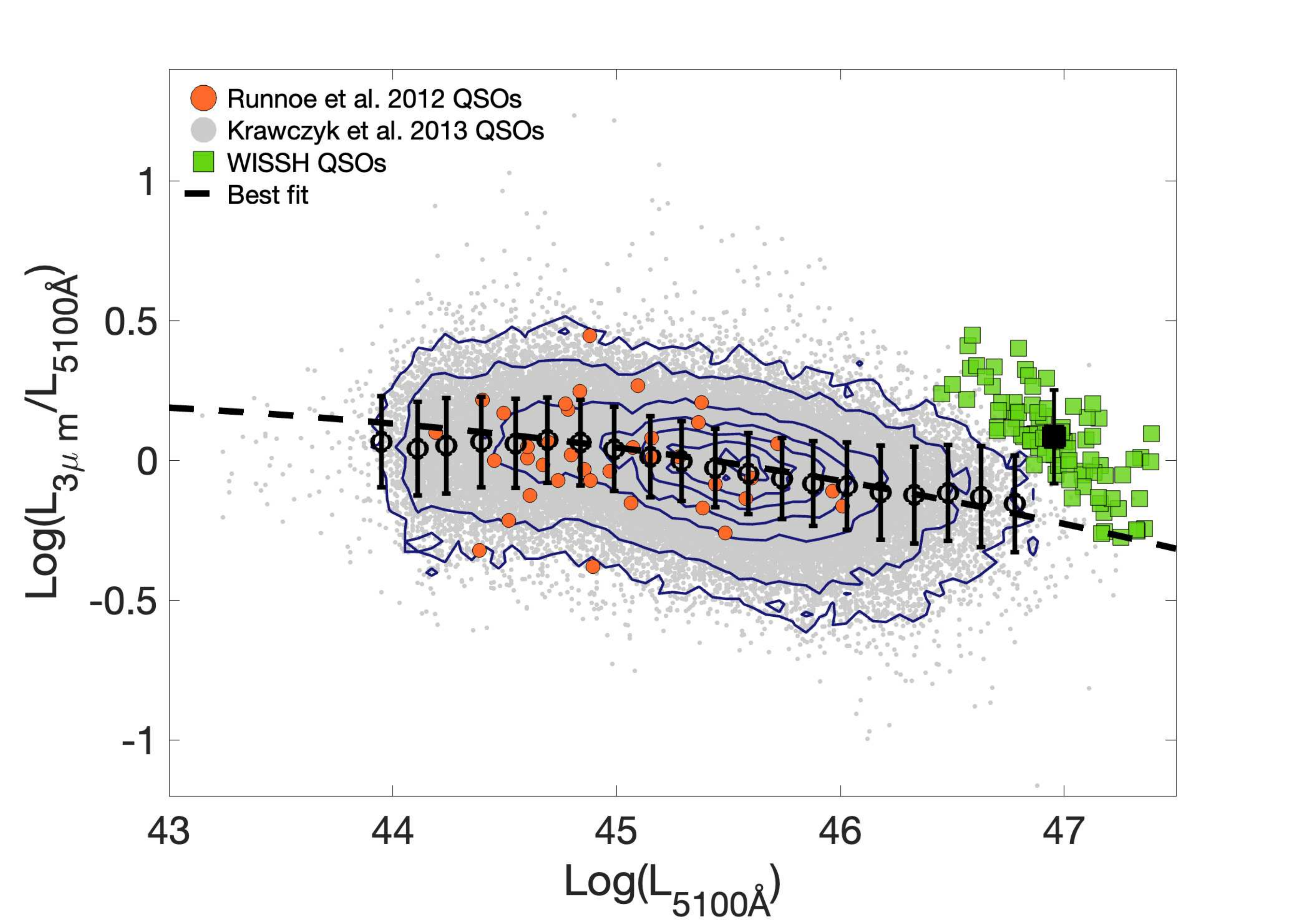}
     \caption{$L_{5100\text{\AA}}$ vs $\log(L_{3\mu m}/L_{5100\text{\AA}})$. QSOs by K13 are included in the plot as grey dots and their median values over bins of width $0.15\,dex$ are portrayed as black circles. The black dashed line gives the best fit obtained assuming the function described in sec. \ref{sec:bolometric_corrections}.  
     WISSH QSOs are depicted as green squares; the black square shows their median value.}
     \label{fig:maiolino}
    
   \end{figure}
In fig. \ref{fig:maiolino} we further examine the evolution of $L_{3\mu m}$ by plotting $\log(L_{3\mu m}/L_{5100\text{\AA}})$ vs $L_{5100\text{\AA}}$. It exists a clear trend showing the $L_{3\mu m}/L_{5100\text{\AA}}$ ratio decreasing as the primary emission (i.e.  $L_{5100\text{\AA}}$) increases with a Spearman correlation coefficient $\rho = 0.37$.
We fitted  K13 data according to the model
\begin{equation*}
    \frac{L_{3\mu m}}{L_{5100\text{\AA}}} = \frac{A}{1+ \left(\frac{L_{5100\text{\AA}}}{10^{45.63}}\right)^{\gamma}}
\end{equation*}
i.e. the same function adopted by \cite{Maiolino:2007},
finding  $A = 1.89\pm 0.01$ and $\gamma =0.25\pm 0.01$. Our fit is plotted as a dashed black line in fig. \ref{fig:maiolino}. We also note a slight change of trend for $L_{5100\text{\AA}} \approx 10^{46.5}$ erg/s, as expected  based on the analysis of the 5100 $\text{\AA}$ and the 3 \textmu m bolometric corrections. However this variation of trend is somewhat less evident than in the case of $K_{3\mu m}$.  

Finally, we note that WISSH QSOs differentiate from the general distribution, being located, on average, 0.3 dex above the best-fit curve.  Moreover, again we observe that the separation from the bulk of the population decreases strongly as luminosity increases, and few brightest sources are in agreement with the fitted curve.



\subsection{BAL vs non-BAL mean SED}
\label{sec:bal_sed}
\begin{figure*}
    \centering
    \captionsetup{width=0.8\linewidth}
    \includegraphics[width=0.8\textwidth]{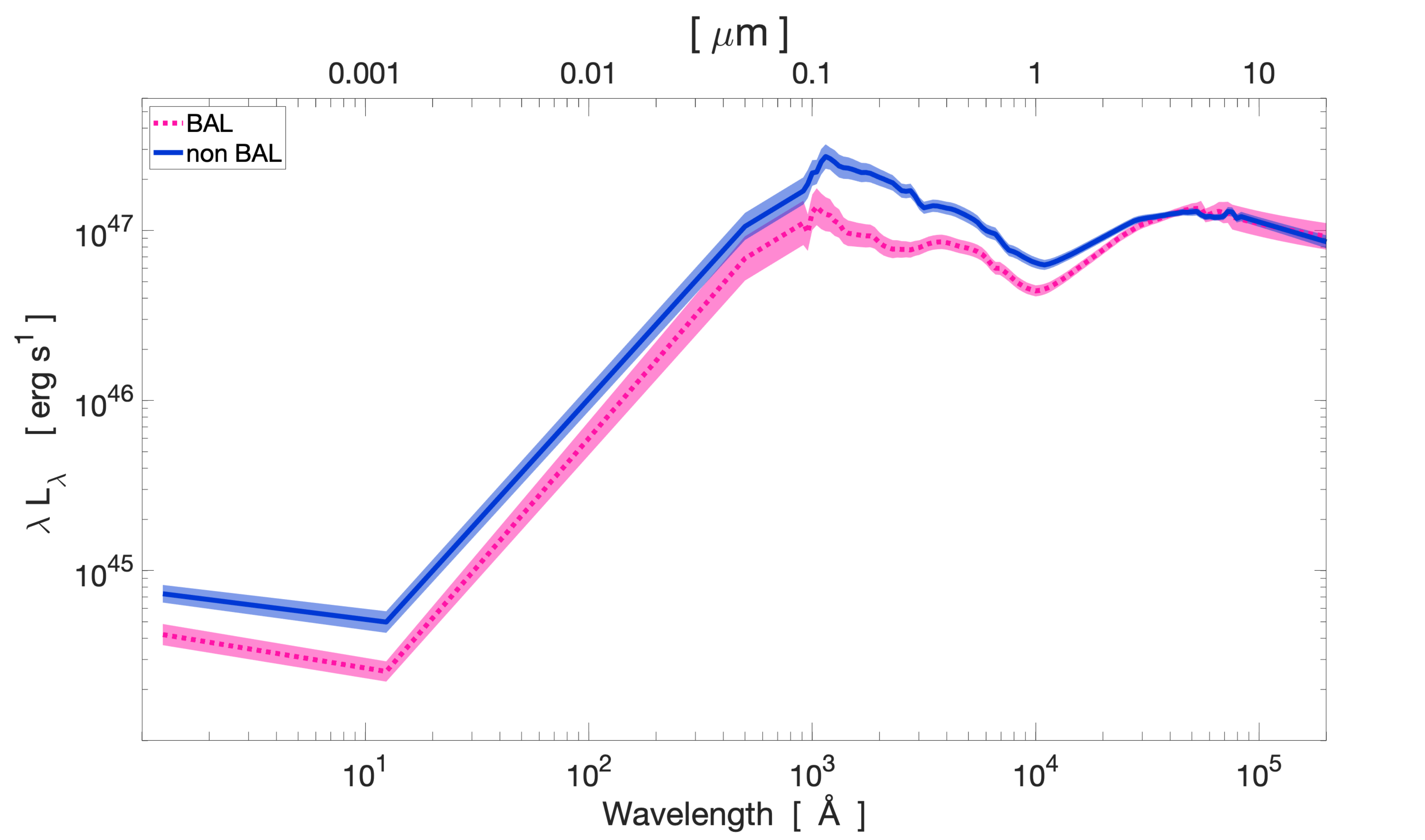}
    \caption{Comparison between the mean SEDs of BAL QSOs (pink dotted line) and non BAL QSOs (blue solid line) together with or 68\% confidence interval shown as shaded area. SEDs are not normalized.}
    \label{fig:sed_BAL}
\end{figure*}
QSOs with Broad Absorption Lines (BALs) in their rest-frame UV spectra represent about 10-20\% of the total AGN population at $z\sim 2-4$ \citep[e.g.][]{Gibson:2009}. These absorption troughs are believed to trace wind on the accretion disk scale \citep[e.g.][]{Hewett:2003}. It is still debated to what extent these winds are present among AGN. A first scenario assumes that all AGN have winds but these are ejected equatorially with a small covering factor  \citep[e.g.][]{Elvis:2000}. In this case objects recognized as BAL are only those whose line of sight intercepts the outflow. Alternatively, BAL winds could be  proper of only a  fraction of the AGN population with peculiar properties, probably associated with a precise AGN evolutionary phase \citep[e.g][]{Wang:2016,Chen:2022}. In any case, whether it is an effect of the viewing angle or to a distinct evolutionary stage, we expect BAL SEDs to differ from that of non-BAL sources. For this reason, we derive  and compare the composite SEDs of BAL and non-BAL subsamples among the WISSH QSOs.\\
For the classification of BAL sources we refer to \cite{Bruni:2019}. In detail, we classify as BAL sources with a modified absorption index $A_{1000} >0$ (34 QSOs). 
In the computation of the mean SED we used only the X-ray luminosity  of the sources observed (13 BAL and 30 non BAL QSOs). This choice allows us to compare only observed fluxes without assuming any correlation between the UV and X-ray regions of the SED. Furthermore, since possible dust extinction is one of the phenomena we are most interested in this analysis, we have not removed red objects with  $\alpha_{opt} < 0.2$ as done in Sec. \ref{sec:removing_red}. However, we have  maintained the \textit{gap repair} performed on photometry with strong absorption features in order to be sure that any differences between the SEDs are not attributable to the overlapping of the filters bandpasses with the absorption troughs.\\
We present the comparison between the mean SEDs of BAL and non-BAL subsamples in fig. \ref{fig:sed_BAL}, while the derived templates are reported in table \ref{tab:SED}.\\
The main differences we notice are in the X-rays  and in the optical and UV region ($1000 \,\text{\AA}<\lambda <1$ \textmu m). In particular, BAL QSOs are X-ray weaker (BALs SED is $\sim0.3$ dex less luminous) and show a redder UV-opt continuum compared to non-BAL sources.\\ 
While the difference in the X-ray seems to be intrinsic, as already found by previous works \citep[e.g][]{Brandt:2000, Gallagher:2002, Gallagher:2007_BAL_SED, Fan:2009, Luo:2014}, 
in the optical and UV band, the redder BAL SED is instead explained as the result of a higher dust extinction \citep[see also][]{Reichard:2003b, Trump:2006, Gallagher:2007_BAL_SED, Dai:2008, Krawczyk:2015,Gaskell:2016}. This is also in agreement with redder WISE W1-W2 colors (rest frame optical bands) in $z\sim6$ BAL quasars with respect to non-BALs found by \cite{Bischetti:2022}.\\

\begin{figure}
    \centering
    \includegraphics[width=0.4\textwidth]{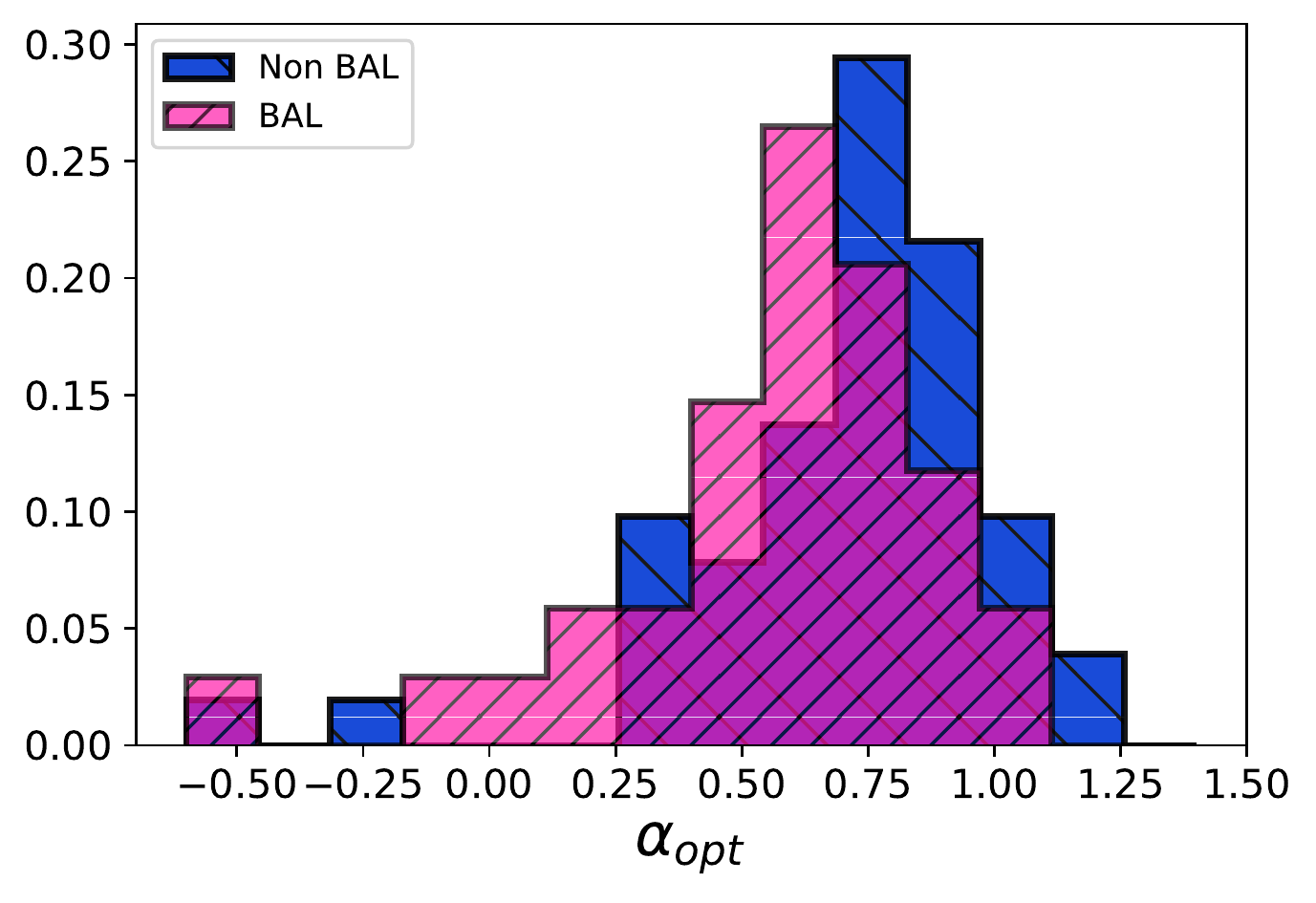}\\
    \includegraphics[width=0.4\textwidth]{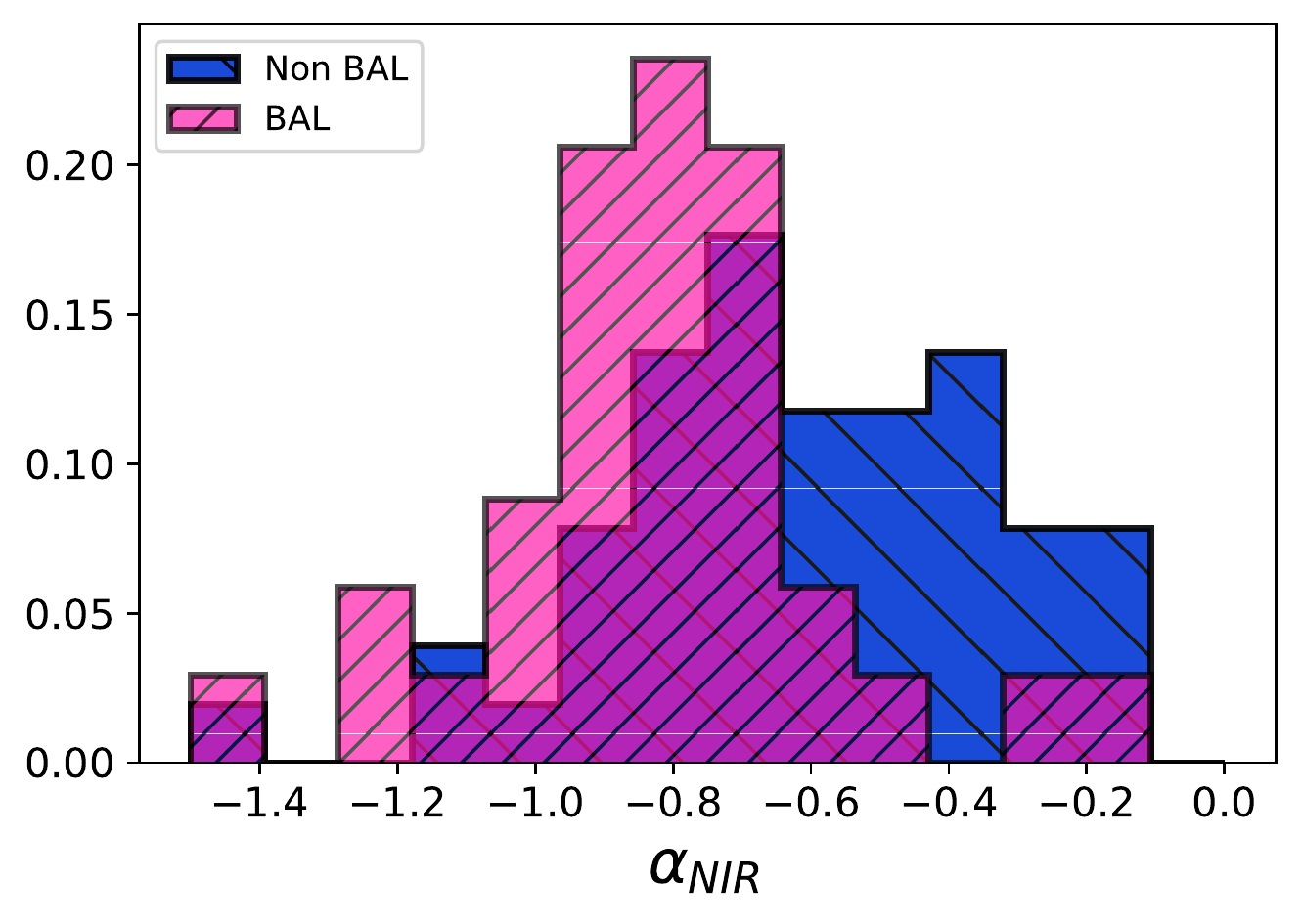}\\
    \caption{\textit{Top Panel:} Normalized histograms of optical spectral slopes between $3000 \,\text{\AA}$ and 1 \textmu m  for BAL (pink) and non BAL (blue) sources. \textit{Bottom Panel:} Normalized histograms of NIR spectral slopes between 1 \textmu m and 4 \textmu m  for BAL (pink) and non BAL (blue) sources.}
    \label{fig:slope_bal}
\end{figure}

Indeed WISSH BAL QSOs show flatter UV emission as visible from the distribution of the optical slope $\alpha_{opt}$ for BAL and non-BAL shown in the top panel of fig. \ref{fig:slope_bal}.
According to the K-S test the two samples are not drawn from the same probability distribution with a p-value of 0.0016. Assuming the same intrinsic emission, this difference can be interpreted entirely as dust extinction. Indeed the average $E_{B-V}$ derived in Sec. \ref{sec:bolometric_lum}, is slightly higher for BAL objects (0.08  compared to 0.03 for non BAL sources) and, furthermore, computing the optical slopes with intrinsic luminosities leads to compatible distributions. \\
Moving to longer wavelengths ($\lambda >1$ \textmu m), 
the differences between the two SEDs are less evident: their emission is similar at $\lambda >3$ \textmu m but BAL SED has a steeper NIR  slope ($1 <\lambda / \mu m < 3$, $\alpha_{NIR}(BAL) = -0.75\pm 0.1$, $\alpha_{NIR}(non BAL) = -0.47 \pm 0.08$), as also visible for individual QSOs in the bottom panel of fig. \ref{fig:slope_bal}.
Given the higher $E_{B-V}$, we would have expected more reprocessed radiation. As an example, assuming a dust covering factor of $4\pi$ and an $E_{B-V} = 0.07$,  we would expect BAL to be $\sim 43\%$ IR brighter compared to non BAL QSOs \citep[see e.g.][]{Gallagher:2007_BAL_SED}. In the case of the analyzed WISSH sample we do not note this IR excess.
However, the higher steepness in the NIR indicates a relatively larger contribution by hot dust.  Performing a K-S test on the distribution of the NIR slopes of BAL and non BAL QSOs we find that the two samples are not drawn from the same distribution with a p-value of 0.0018. \\
Similarly \cite{Zhang:2014} found a small but significant correlation between the NIR slope and the BAL parameters such as the blue-shifted velocity and the balnicity index. To explain the existence of these relationships, they suggested the presence of a dust component  within the outflowing gas. This dust can be  co-spatial with the gas-clouds and therefore be originally mixed with the outflow or it could be intercepted by the outflow once it interacts with the innermost regions of the torus. In both cases, the effect is a steepening of the NIR slope. In the first case this is due simply because  outflows are believed to originate very close to the accretion disk and the dust grains, there contained, could heat up to  very high temperatures. In the latter case, hydro-dynamical simulations showed that the effect of the  powerful wind impacting  dense clouds is to break them into  diffuse warm filaments. In this way more dust is directly exposed to the central UV source and therefore heated to higher temperature.

\subsection{Mean SEDs based on $CIV\, REW$}
\begin{figure*}
    \centering
    \captionsetup{width=0.8\linewidth}
    \includegraphics[width=0.8\textwidth]{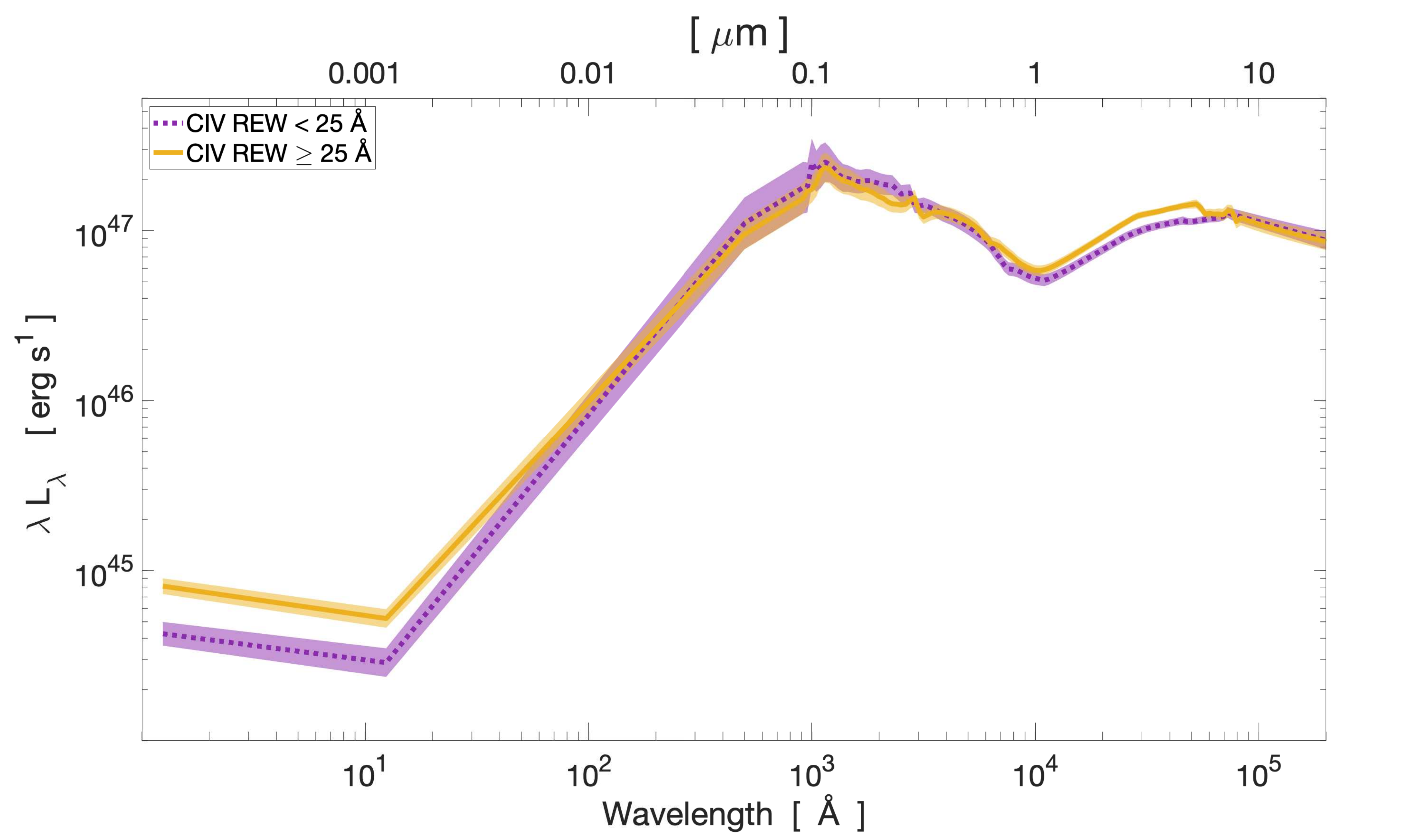}
    \caption{Comparison between the mean SEDs of QSOs with $REW<25\,\text{\AA}$ (purple dotted line) and those with $REW\geq 25\,\text{\AA}$ (yellow solid line) together with or 68\% confidence interval shown as shaded area. SEDs are not normalized.}
    \label{fig:CIV_SED}
\end{figure*}
\begin{figure}
    \centering
    \includegraphics[width =0.4\textwidth]{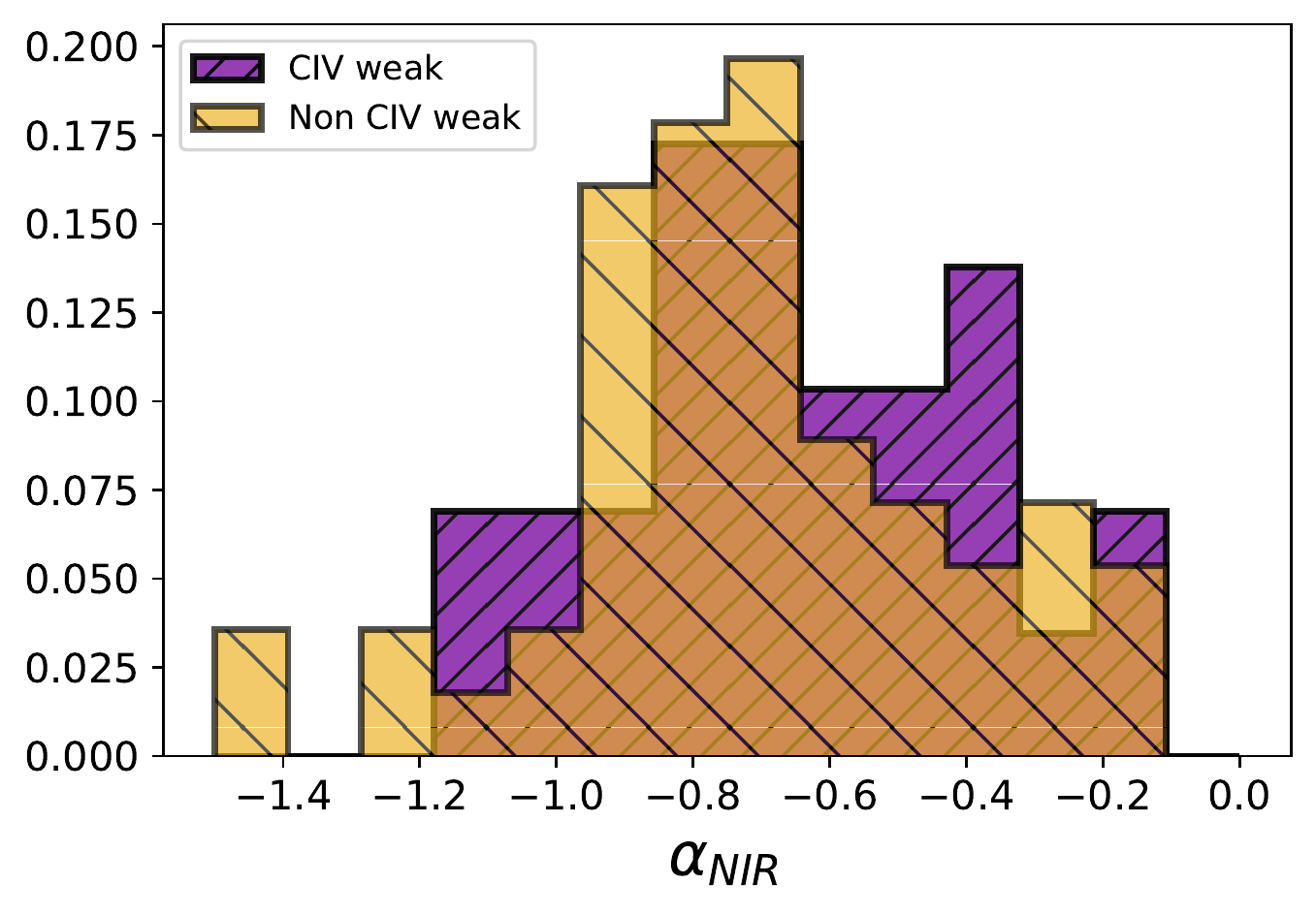}
    \caption{Normalized distributions of NIR spectral slopes $\alpha_{NIR}$ for QSOs with $REW_{CIV}<25\, \text{\AA}$ (purple) and those with $REW_{CIV}\geq25\, \text{\AA}$ (yellow).}
    \label{fig:beta_civ}
\end{figure}

The CIV line at $1549\,\text{\AA}$, like other high ionization lines such as SiIV and [OIII], can show a broad asymmetric profile and be blue-shifted with respect to the systemic redshift; this can be explained by assuming that the emitting region is moving toward the observer. For this reason, this line is widely studied as a tracer of outflows on the Broad Line Region scale. Moreover it is observed that CIV line and especially its relative strength, expressed through the rest frame equivalent width (\textit{REW}), is linked to  several of the AGN properties. Probably the best known relationship is the Baldwin effect, 
i.e the anti-correlation between the $1550 \, \text{\AA}$ luminosity and the CIV $REW$ \citep[][]{Baldwin:1977}. Firstly observed on a sample of 20 sources, this relation has been confirmed  with significantly larger samples \citep[e.g][]{Wu:2009, Richards:2011} although some authors \citep[][]{Baskin:2004,Shemmer:2015} suggest that this is a secondary effect and that $REW_{CIV}$ primarily (anti)correlates with the Eddington ratio. Under this hypothesis, a larger $L/L_{EDD}$ value would produce a more UV-peaked SED \citep[][]{Shemmer:2015}, decreasing the number of ionizing photons.\\
To test and analyze any possible difference between objects with different $REW_{CIV}$, we derived the 
composite SEDs by dividing WISSH QSOs according to their $REW_{CIV}$. 
We tried to make the sub-sample with lower $REW$ to be  representative of the Weak Lines Quasars \citep[WLQs,][]{Fan:1999,Diamond-Stanic:2009a}, a population of objects that shows the highest measured blue-shifts of the lines and is therefore associated to the fastest winds \citep[e.g.][]{Wu:2011,Luo:2015}.  
WLQs  are usually defined by having $REW <10 \,\text{\AA}$  \citep[e.g][]{Shemmer:2009}. According to \cite{Shen:2011}, to whom we refer for the REW measures, only three WISSH quasars  satisfy this criterion and can be considered as WLQS. However, \cite{Vietri:2018} found that WISSH QSOs with $REW_{CIV} < 20 \,\text{\AA}$ show outflow velocities comparable with those of proper weak line emitters, suggesting that in  case of particularly bright quasars, the threshold could be extended to include also these sources.\\ 
In the construction of the composite SEDs based on \textit{CIV} $REW$, we softened even more the selection criteria to have two sub-samples both with a sufficient number of elements, i.e. 29 QSOs with $REW_{CIV} \leq 25 \,\text{\AA}$ and 56 with $REW_{CIV} > 25 \,\text{\AA}$ (instead of 17 QSOs with $REW_{CIV} \leq 20 \,\text{\AA}$ and 68 with $REW_{CIV} > 20 \,\text{\AA}$).
Although quasars  with larger blueshifts corresponding to  smaller $REW$ are usually found to have deeper absorption troughs \citep[][]{Rankine:2020}, we note that in our sample this is not generally valid; indeed only 6 sources are classified  as both BAL and weaker CIV emitters ($\sim 21\%$ of BALs).\\
To derive the mean SEDs we used the photometry as described in Sec. \ref{sec:bal_sed}. 
The comparison between the SEDs is presented in fig. \ref{fig:CIV_SED} and the derived templates are provided in table \ref{tab:SED}.\\
The main differences between the two SEDs lie in the X-ray region where QSOs with a weaker CIV line have a lower emission of about 0.4 dex. This result is in agreement with several previous works (e.g.\citealt{Wu:2009, Richards:2011, Kruczek:2011,Timlin:2020, Zappacosta:2020} but see also \citealt{Lusso:2021}) and is also expressed in terms of a steeper $\alpha_{OX}$. 
In the UV part of  our mean SEDs, we do not find a strong evidence for the Baldwin effect, the CIV 'weak' composite SED is indeed slightly more luminous in the UV but the SEDs are compatible within the uncertainties. Performing a Spearman correlation test on the intrinsic $L_{1550\text{\AA}}$ and $REW_{CIV}$ for each individual QSO returns the null hypothesis probability (i.e. no correlation exists) with a p-value of 0.47. This result suggests that the lack of evidence of the Baldwin effect in the mean SEDs does not depend on the chosen $REW_{CIV}$ cut  for splitting sample but likely rather on the fact that we are studying sources already at the bright end of optical luminosity function and thus little variance among them is expected \citep[see also the discussion in Sec. 4.4 from][]{Lusso:2021}. \\ 
Moving to the IR region, we find no significant difference between the SEDs. This is in contrast with a recent work by \cite{Temple:2021} \citep[but see also][]{Wang:2013} who, analyzing a sample of $\sim 5000$ QSOs at $z\approx2$ and $L_{bol}\sim 10^{47}$ erg/s, found a negative correlation between  the NIR spectral slope (giving the hot dust to nuclear emission ratio) and the CIV $REW$. 
Even investigating the spectral slopes individually (fig. \ref{fig:beta_civ}), we do not find evidence in favor of a difference between the two distributions. A K-S test returns indeed the null hypothesis  with a p-value of 0.83.

\begin{table*}[ht]
\begin{center}
\begin{tabular}{ccccccccccc}

$\log(\lambda )$ & All &$\sigma $& BAL & $\sigma $& non-BAL &$\sigma $ & $REW_{CIV}<$ 25 \AA &$\sigma $& $REW_{CIV}\geq$ 25 \AA &$\sigma $\\
\hline
\hline
0.08 & 44.92 & 0.03 & 44.63 & 0.06 & 44.87 & 0.05 & 44.63 & 0.07 & 44.63 & 0.07\\
0.10 & 44.92 & 0.03 & 44.62 & 0.06 & 44.86 & 0.05 & 44.63 & 0.07 & 44.63 & 0.07\\
0.12 & 44.91 & 0.03 & 44.62 & 0.06 & 44.86 & 0.05 & 44.62 & 0.07 & 44.62 & 0.07\\
0.14 & 44.91 & 0.03 & 44.61 & 0.06 & 44.86 & 0.05 & 44.62 & 0.07 & 44.62 & 0.07\\
0.16 & 44.91 & 0.03 & 44.61 & 0.06 & 44.85 & 0.05 & 44.62 & 0.07 & 44.62 & 0.07\\
0.18 & 44.90 & 0.03 & 44.6 & 0.06 & 44.85 & 0.05 & 44.61 & 0.07 & 44.61 & 0.07\\
\hline
\end{tabular}
\end{center}

\caption{Mean SEDs computed for the WISSH full sample and for the subsamples: BAL, non-BAL, $REW_{CIV}<$ 25 \AA, $REW_{CIV}\geq$ 25 \AA. Units are $\lambda$ [\AA] and $\lambda L_{\lambda} \; [erg\,s^{-1}]$.\\
The full table is available in the online version.}
\label{tab:SED}
\end{table*}

\subsection{SF versus AGN dominated sources}
\begin{figure*}
    \centering
    \includegraphics[width =0.9\textwidth]{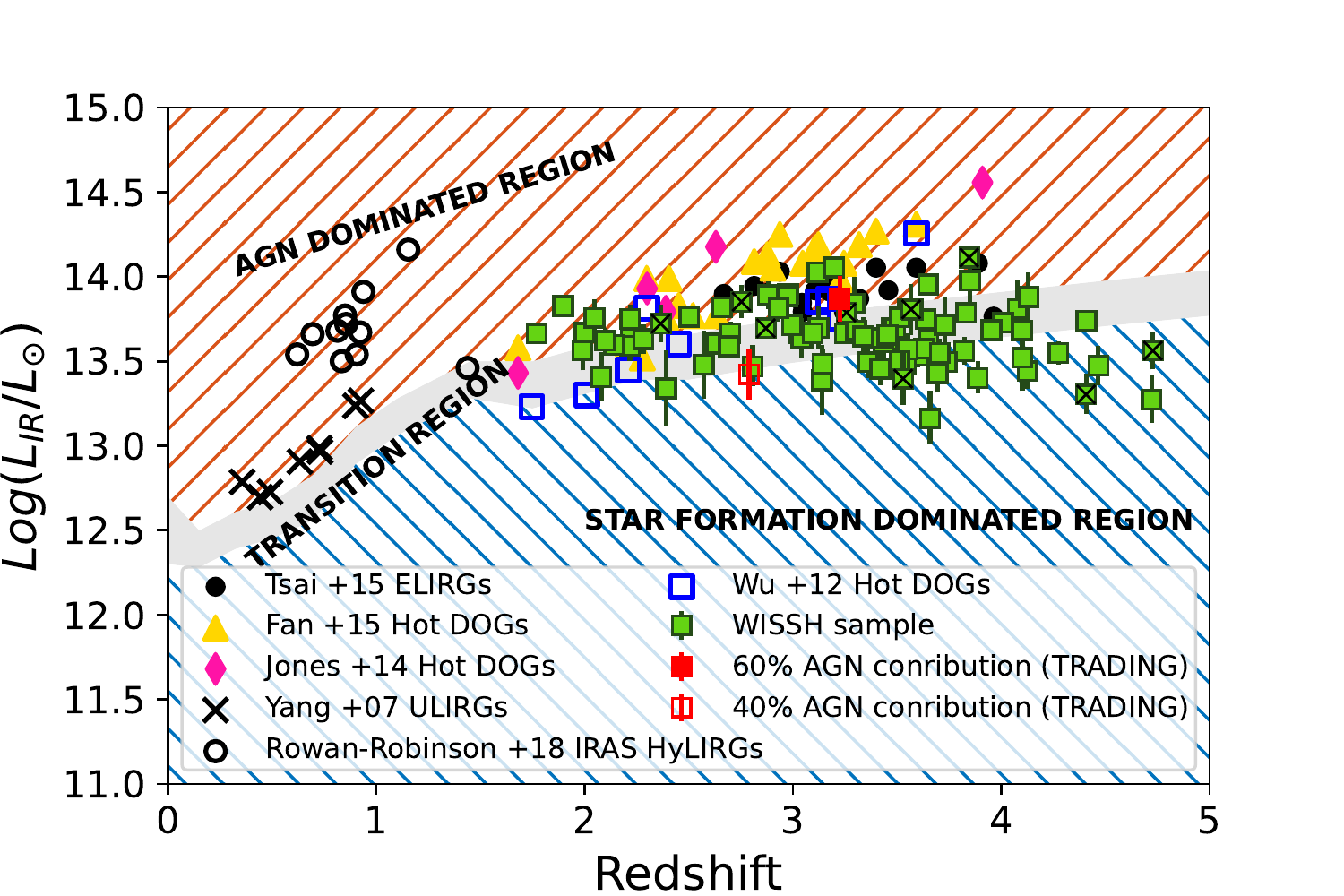}
    \caption{[Adapted from fig. 8 of \cite{Symeonidis:2021}] Redshift vs $L_{8-1000 \, \mu m}$ diagram partitioned into an  AGN dominated, a transition and a star formation dominated region. Overplotted are various samples from the literature:  the \textit{IRAS}-selected HyLIRGs from \cite{Rowan-Robinson:2018}, the intermediate redshift ULIRGs from \cite{Yang:2007}, the optically unobscured QSOs from \cite{Tsai:2015} and the hot dust obscured galaxies (hot DOGs) from \cite{Fan:2016}, \cite{Jones:2014} and \cite{Wu:2012}. The WISSH sample studied in this work is shown as green squares. Markers with an 'x' inside represent QSOs with FIR \textit{ALMA/NOEMA} coverage. Plotted as red squares are WISSH08 and WISSH51 for which \cite{Duras:2017} evaluated the AGN contribution to $L_{IR}$.}
    \label{fig:symeonidis_plot}
\end{figure*}
In a recent work, \cite{Symeonidis:2021} studied the evolution with redshift of the fraction of AGN dominated sources for a given value of the IR $8-1000$ \textmu m integrated luminosity $L_{IR}$. Assuming that galaxies are totally AGN powered or SF powered, this fraction is given by $\mathcal{F}$, defined as the ratio between the AGN IR Luminosity Function $\phi_{AGN}$, derived from the X-ray LF by \cite{Aird:2015}, and the galaxy IR luminosity function $\phi_{gal}$ by \citet{Gruppioni:2013}.\\
Defining as $L_{25}(z)$ and $L_{75}(z)$ the luminosity values for which  $\mathcal{F}$ is respectively 0.25 and 0.75, \cite{Symeonidis:2021} defined three regions in the $z-L_{IR}$  space: the AGN dominated region for sources with $L_{IR} > L_{75}$, the transition region for those with $L_{25} \leq L_{IR} \leq L_{75}$ and the star formation dominated  region for galaxies with $ L_{IR} < L_{25}$. As it can be seen in fig. \ref{fig:symeonidis_plot}, the IR luminosity required to be in the AGN dominated region increases with z.   
Moreover, \cite{Symeonidis:2021} filled this $z-L_{IR}$ diagram with quasar samples taken from various catalogues (including about half of WISSH QSOs), each claiming to contain among the brightest QSOs for a given range of redshift. Being very luminous, almost all these sources are located in the AGN dominated region (see fig. 8 from their work).

We added to this diagram the whole WISSH sample. WISSH $L_{IR}$ have been derived using the same procedure discussed in Sec. \ref{sec:bolometric_lum} but in this case we limited the fit to \textit{WISE, Herschel} and \textit{ALMA} or \textit{NOEMA} photometry and fixed the $E_{B-V}$ to the already computed value. 
We expect to get the more accurate $L_{IR}$ estimates for sources with measured \textit{ALMA/NOEMA} fluxes; indeed, these observations, in addition to providing a better sampling of the integration interval, are high-resolution and spatially resolved, i.e. the measured emission comes exclusively from the active galaxy and is not affected by the possible presence of a nearby companion.\\ 
When included in the $z-L_{IR}$ diagram, almost all WISSH sources ($\sim 88\%$) up to $z\sim 3.5$ are in agreement with the proposed partition, being either in the AGN dominated or in the transition region  (see fig. \ref{fig:symeonidis_plot}).
However, this percentage drops to $\sim 50\%$ at $z>3.5$. 
Although, as clearly stated by \cite{Symeonidis:2021}, this $z-L_{IR}$ diagram does not have the pretension to be a precise diagnostic instrument and it could happen that AGN dominated sources fall in the starbust region and vice versa, it would seem that, despite their high luminosities, the FIR emission of WISSH QSOs at $z\sim 3.5$ is still mostly powered by star formation activity. This conclusion is supported by \cite{Bischetti:2021} who found WISSH QSOs to be located in in high star-forming environments. On the other hand, our findings disagree with those by \cite{Symeonidis:2022} who determined that, for QSOs with $L_{5100\text{\AA}} \gtrsim 10^{46}$ erg/s, $L_{IR}$ essentially traces the AGN primary emission and that the contribution provided by SF becomes almost negligible. However, although their results in the high-luminosity regime seem to be redshift independent, the sample studied by \cite{Symeonidis:2022} consists only of QSOs up to $z = 2.65$. 
Furthermore, our conclusions depend (mostly) on the diagram partition. It should be remembered that
in the original $z-L_{IR}$ diagram, $\mathcal{F}$ was computed by \citet{Symeonidis:2021} only up to $z=2.5$  and then extrapolated up to $z=4$   based on the evolution of $\phi_{gal}$ (and we extrapolated up to $z=5$ to include all WISSH QSOs, moving further away from the region constrained by obsevations). The reason behind this choice is the severe lack of spectroscopic redshifts amongst the population that makes up the IR LF at $z > 2.5$ \citep{Symeonidis:2021}. As a result, the possibility that our findings are influenced by the fact that we are looking at a portion of the diagram where the partition is rather uncertain cannot be ruled out. 
As a partially independent test for the proposed partition, we took into account results by \cite{Duras:2017} who, using the radiative transfer code TRADING  \citep[][]{Bianchi:2008} on WISSH08 and WISSH51 (i.e. the two QSOs with the highest and lowest luminosity as measured by \textit{Herschel}), established the AGN contribution to $L_{IR}$ to be respectively 60\% and 40\%. By looking at the placement of these two QSOs in the diagram, we find that they are located in the transition region and thus in agreement with the independently computed AGN contribution. Unfortunately, these sources have $z=3.23 $ and $z=2.89$ and so they only marginally test the high redshift region, which is the one we are most interested in.  

\section{Summary and conclusions}
In this work we derived the mean SED of the WISSH sample: 85 hyper-luminous type 1 QSOs. Our results can be summarized as follows:
\begin{itemize}
    \item Overall the shape of the mean SED of these sources is very similar to that of the bulk of type 1 quasar population with the main differences being observed in the X-ray region where WISSH quasars are relatively weaker and in the near and mid IR bands  where instead they show an excess in their emission. This excess is responsible for both a more prominent red bump and for the shifting of the SED dip from 1.3 to 1.1 \textmu m.
    \item The IR excess has been previously reported in literature and can be explained by assuming an extra contribution from  dust. We find that to properly model this IR excess two distinct dust components are required: 1) warm dust ($T \approx 450-800$ K), probably associated to the torus and responsible for the enhancement of the IR bump longwards than $\sim 3$ \textmu m and 2) hot dust close to the sublimation temperature ($T > 1000$ K) which accounts for the divergences at $\lambda \sim 1-3$ \textmu m. 
    Further investigations are needed to confirm both the shift of the dip and the origin of this extra hot dust contribution that we are unable to explain in the context of classical AGN models.
    \item The relatively lower X-ray emission is an already well known feature of luminous QSOs and it is in agreement with the $\alpha_{OX}-l_{UV}$ anti-correlation \citep[e.g.][]{Vignali:2003,Lusso:2010,Martocchia:2017} . 
    \item By modelling the QSOs emission with the mean SED and integrating it between $10$ keV and $1$ \textmu m, we computed the bolometric luminosities, confirming that WISSH quasars are indeed among the most luminous AGN, having all $L_{bol} > 10^{47}$ erg/s. By combining $L_{bol}$ with the monochromatic luminosities, we then derived the  $5100\,\text{\AA}$ and 3 \textmu m bolometric corrections. We compared our results with data by \cite{Runnoe:2012} and K13. In  the case of $K_{5100\text{\AA}}$, WISSH QSOs have a similar distribution to that of lower luminosity QSOs and are in agreement with both relationships proposed by \cite{Runnoe:2012} and K13. For $\lambda = 3$ \textmu m we observe in WISSH QSOs a lower bolometric correction than in the bulk of the population; 
    We also derive a new luminosity-dependent $K_{3\mu m}$ described by a piecewise function, i.e. with an increasing trend up to $L_{bol} \approx 10^{47}$ erg/s, followed by a slightly decreasing one at higher luminosities. 
    \item An increasing $K_{3\mu m}$ vs $L_{bol}$ is in agreement with
     \cite{Maiolino:2007} who interpreted such trend as due to a  decrease of torus covering factor with increasing luminosities. However, the observed high luminosity change of $K_{3\mu m}$ trend might indicate a limiting radius beyond which an increase in the primary emission does not correspond to a further receding of the torus but rather to an increment of dust heated up to its sublimation temperature. This would be in agreement with what is found in WISSH mean SED.
\end{itemize}
We also derived the mean SEDs by splitting the sample according to their spectral features: BAL vs non-BAL and $REW_{CIV} \leq 25\text{\AA}$  vs $REW_{CIV} >25\text{\AA}$. We find that:
\begin{itemize}
    \item BALs exhibits  lower X-ray emission  and a depressed UV to optical continuum. The X-ray weakness is usually considered intrinsic since it is present even when luminosities are corrected for the absorption; the flatter UV-optical spectrum is  instead compatible with a larger extinction by dust. We also note that BALs have a steeper NIR slope which indicates a higher contribution by the hottest  dust component, this result is in agreement with  the analysis by \cite{Zhang:2014}.
    \item We find a clear dichotomy between the CIV 'weak' and 'non-weak' populations regarding their X-ray emission.  QSOs with a smaller equivalent width have an intrinsic lower X-ray emission. This result is consistent with the already known $REW_{CIV}$ - X-ray correlation. As in the case of BALs, the lower X-ray output from weaker CIV emitters seems to be intrinsic. For our quasars the Baldwin effect is not so evident to appreciate its effect neither in the comparison between the mean SEDs nor investigating the individual quasar SEDs. We also do not find any evidence for the $REW_{CIV}$ and the NIR spectral  slope correlation which was recently suggested by \cite{Temple:2021}.
\end{itemize}
Finally we included the WISSH quasars in the $z-L_{IR}$ diagram by  \cite{Symeonidis:2021} which explores the evolution with redshift of the fraction of AGN dominated vs SF dominated sources. We find that up 
to z $\sim$ 3.5, $\sim88\%$ of WISSH QSOs fall either in the AGN-dominated or transition region, while this percentage drops below 50\% for higher redshifts. This last finding would suggest that even QSOs specifically selected to be the most luminous have FIR emission mostly dominated by SF activity.  In any case, this result relies on the strong and non trivial assumption that the partition proposed by \cite{Symeonidis:2021} still holds for z > 3.5, which we cannot prove and therefore we are unable to draw any firm conclusions.



Concluding, the mean SED of extremely luminous WISSH QSOs exhibits non-negligible differences compared to that of less luminous sources. Those differences are likely present also in other samples of hyper-luminous quasars such as, for instance, those in the  high-z Universe  \citep[z>6.5, e.g.][]{Mazzucchelli:2017}. For this reason, in the analysis of the latter we recommend to use SED templates specifically constructed from very luminous QSOs. \\

\begin{acknowledgements}
We are grateful to the anonymous referee for their useful comments and suggestions which helped us to improve the paper.
I.S. thanks Coleman. M. Krawczyk for providing helpful advice on the derivation of the mean SED. I.S. also thanks Myrto Symeonidis for providing $L_{IR}$ data.
M.B. and E.P. acknowledge support from PRIN MIUR project "Black Hole winds and the Baryon Life Cycle of Galaxies: the stone-guest at the galaxy evolution supper", contract \#2017PH3WAT. R.M. acknowledges support by the Science and Technology Facilities Council (STFC) and ERC Advanced Grant 695671 "QUENCH". R.M. also acknowledges funding from a research professorship from the Royal Society.
\end{acknowledgements}

\bibliographystyle{aa}
\bibliography{ivano}

\appendix
\section{TNG data reduction}
\label{sec:TNG_data}
Near-IR data have been collected during  Program A37TAC\_32 in period 37 and program A42TAC\_17 in period 42; PI: V. Testa.
The reduction has been performed following a standard procedure currently adopted for
this type of detectors: images were acquired adopting a dithering strategy obtaining nine or ten image per cycle, for each filter. These were then median stacked using a clipping algorithm to obtain an empty-sky frame, that has then been subtracted to all the single frames. The output of this stage is a set of images with a zero counts background level. This procedure removes bias and dark current levels together with the sky, and provides a zeroth order flat fielding. The following step was to normalize all the single frames with a flat-field image obtained using twilight sky frames as suggested on the telescope site. The flat-fielded images were then registered and co-added obtaining the final frames for each filter, on which the data analysis has been performed. We used a set of semi-automatic procedures using IRAF and DAOPHOT and the instrumental magnitudes were calibrated through a set of stars in the field with the 2MASS survey.


%

%

\end{document}